\documentclass[a4paper,11pt]{article}
\pdfoutput=1 
\usepackage{jheppub} 
\usepackage[T1]{fontenc} 
\usepackage{natbib}
\usepackage{youngtab}
\usepackage{hyperref}
\usepackage{fancybox}
\usepackage{float}
\usepackage{amsfonts}
\usepackage{amsmath}
\usepackage{amsbsy}
\usepackage{graphicx}
\usepackage{amssymb}
\usepackage{amsthm}
\usepackage{bm}
\usepackage{comment}
\usepackage{epsfig}
\usepackage{latexsym}
\usepackage{color}
\usepackage{slashed}
\usepackage{physics}
\usepackage{mathtools}
\usepackage{scalerel}
\usepackage[usestackEOL]{stackengine}
\usepackage{dsfont}
\usepackage{footmisc}
\usepackage{subcaption}

\renewcommand{\tilde}{\widetilde}

\renewcommand{\Im}{{\rm Im} \,}

\DeclareMathOperator{\sn}{sn}
\DeclareMathOperator{\cn}{cn}
\DeclareMathOperator{\dn}{dn}

\DeclareMathOperator{\supp}{supp}

\newcommand{\beq}{\begin{eqnarray}}
\newcommand{\eeq}{\end{eqnarray}}
\newcommand{\bea}{\begin{eqnarray}}
\newcommand{\eea}{\end{eqnarray}}
\newcommand{\be}{\begin{equation}}
\newcommand{\ee}{\end{equation}}
\newcommand{\bq}{\begin{equation}}
\newcommand{\eq}{\end{equation}}

\newcommand{\nn}{\nonumber}

\def\ie{\begin{equation}\begin{aligned}}
\def\fe{\end{aligned}\end{equation}}
\def\ieg{\begin{equation}\begin{gathered}}
\def\feg{\end{gathered}\end{equation}}

\newcommand{\AdS}{{\rm AdS}}
\newcommand{\EAdS}{{\rm EAdS}}
\newcommand{\adsS}{{\rm AdS}_5 \times S^5}

\newcommand{\pa}{\partial}
\def\Tr{\text{Tr}}
\newcommand{\la}{\langle}
\newcommand{\ra}{\rangle}

\newcommand{\cA}{{\mathcal A}}

\newcommand{\cN}{{\mathcal N}}
\newcommand{\cO}{{\mathcal O}}

\def\cO{{\cal O}}

\def\n{\nu}

\def\6{\partial}

\def\tr{{\rm Tr}}

\def\la{\langle}
\def\ra{\rangle}

\def\dashint{\,\ThisStyle{\ensurestackMath{%
            \stackinset{c}{.2\LMpt}{c}{.5\LMpt}{\SavedStyle-}{\SavedStyle\phantom{\int}}}%
        \setbox0=\hbox{$\SavedStyle\int\,$}\kern-\wd0}\int}

\def\6{\partial}

\title{Bubbling wormholes and matrix models} 

\author[a]{Panos Betzios,}
\author[b]{Ji Hoon Lee,}
\author[c]{Olga Papadoulaki,}
\author[d]{Yanjun Zhou}
\affiliation[a]{\href{https://www.ugent.be/we/physics-astronomy/en}{Department of Physics and Astronomy},
Ghent University, \\ Krijgslaan, 281-S9, 9000 Gent, Belgium}
\affiliation[b]{Institut f\"{u}r Theoretische Physik, ETH Zurich, CH-8093 Z\"urich, Switzerland}
\affiliation[c,d]{\href{https://www.cpht.polytechnique.fr/}{CPHT}, CNRS, \'Ecole polytechnique, Institut Polytechnique de Paris, 91120 Palaiseau, France}
\emailAdd{panos.betzios@ugent.be} \emailAdd{jileej@phys.ethz.ch}\emailAdd{olga.papadoulaki@polytechnique.edu}\emailAdd{yanjunzhou2001@gmail.com}

\abstract{The thermofield double state entangles two copies of a CFT via a sum over energy eigenstates and is dual to the two-sided eternal black hole. We explore an analogous construction using sums over gauge group representations of half-BPS Wilson loops in multiple copies of $U(N)$ $\mathcal{N}=4$ super Yang-Mills. These sums act as delta function-like operators that correlate the eigenvalues of the corresponding half-BPS matrix models. We suggest that the holographic duals are ``bubbling wormhole'' geometries: multi-covers of AdS$_5 \times S^5$ whose conformal boundary consists of multiple four-spheres intersecting on a common circle. We analyze the matrix model free energy, discuss its bulk interpretation, and study probe loops in these backgrounds.}

\begin{document} 
\maketitle
\flushbottom

\section{Introduction}

A central theme in holography is the connection between entanglement and geometry. The thermofield double state, which entangles two copies of a CFT via a sum over energy eigenstates, is dual to the two-sided eternal black hole, a connected geometry with two asymptotic boundaries \cite{Maldacena:2001kr}. This work explores an analogous construction in which entanglement is considered over gauge group representations $R$ rather than energy eigenstates\footnote{Similar entangled sums over representations, have been considered in~\cite{Betzios:2021fnm,Betzios:2023obs} in relation to \emph{Euclidean} wormhole geometries.}.

A well-studied observable in $\cN = 4$ super Yang-Mills is the circular half-BPS Wilson loop
\be\label{def. of half-BPS Wilson loop in N=4 SYM}
W_R = \tr_R P \exp \left[ \oint d s \, ( i A_\mu \dot{x}^\mu + \Phi_0 )  \right] \, ,
\ee
transforming in the irreducible representation $R$ \cite{Maldacena:1998im,Rey:1998ik,Drukker:1999zq,Zarembo:2002an}. Via supersymmetric localization \cite{Pestun:2007rz}, its expectation value reduces to a Gaussian matrix model integral, where $W_R$ becomes $\Tr_R(e^M)$ for a Hermitian matrix $M$ \cite{Erickson:2000af,Drukker:2000rr,Semenoff:2001xp}. The matrix $e^M$ can be viewed as the holonomy of a complexified gauge connection $A_\mathbb{C}$ around a great $S^1$ of the $S^4$ on which the field theory is placed. The holographic dual of $W_R$ depends on the size of the representation. Wilson loops with $O(1)$ boxes are dual to fundamental strings \cite{Maldacena:1998im,Rey:1998ik,Drukker:1999zq,Zarembo:2002an}, those with $O(N)$ boxes to D-branes \cite{Drukker:2005kx,Hartnoll:2006is,Yamaguchi:2006tq,Gomis:2006sb,Gomis:2006im,Drukker:2007dw,Gomis:2008qa,Faraggi:2011bb,Fiol:2013hna,Zarembo:2016bbk}, and those with $O(N^2)$ boxes to backreacted ``bubbling'' geometries \cite{Yamaguchi:2006te,Lunin:2006xr,DHoker:2007mci,Okuda:2007kh,Okuda:2008px,Benichou:2011aa,Aguilera-Damia:2017znn}.

Consider two copies of $U(N)$ $\mathcal{N}=4$ super Yang-Mills theories SYM$_1$ and SYM$_2$, each defined on $S_1^4$ and $S_2^4$, respectively.\footnote{For simplicity, we restrict the discussion in the introduction to the case with two copies.} While the theories SYM$_{1,2}$ are decoupled, we can consider the insertion of an ``entangled'' sum of pairs of Wilson loops over their irreps
\be\label{correlatedloops}
\sum_R \, \langle W_R \rangle_1 \, \langle W_R \rangle_2 \, ,
\ee  
where $\langle W_R \rangle_A$ is the normalized expectation value in the corresponding copy SYM$_A$.

In this work, we study the expectation values (and the dual supergravity descriptions) of half-BPS specializations of operators \eqref{correlatedloops} such as
\be \label{heavydeterminant}
\sum_R \tr_R(e^{M_1}) \tr_R(e^{-M_2}) = \frac{1}{\det \left(\mathds{1} \otimes \mathds{1} - e^{M_1} \otimes e^{-M_2}  \right)} ,
\ee
where the determinant is taken over the $N^2 \times N^2$ tensor product of two $N \times N$ matrices $M_1$ and $M_2$. While each summand in the expectation value $\la \sum_R \tr_R(e^{M_1}) \tr_R(e^{-M_2}) \ra_{12}$ is factorized into those evaluated in theories SYM$_1$ and SYM$_2$ individually, the sum over all irreps $R$ has a sizable effect on the saddle points of the two half-BPS matrix models. We find that operators such as \eqref{heavydeterminant}, when inserted into the path integrals of SYM$_A$ on $S_A^4$, can have the effect of gluing the various $S_A^4$ along a common great $S^1$ on which the operator is placed. We suggest that the holographic dual of this configuration is a \textit{bubbling wormhole}, a geometry that is a multi-cover of $\AdS_5 \times S^5$, with multiple asymptotic regions whose conformal boundaries $S_A^4$ share a common $S^1$.

That the operator \eqref{heavydeterminant} plays the role of a delta function can be precisely shown in the context of unitary matrices. Consider a pair of unitary matrices $U,V \in U(N)$. The delta function on the $U(N)$ group manifold can be expanded in terms of group characters as
\be\label{deltaexpansionchar}
\delta(U ,V) = \sum_R \chi_R(U) \chi_{R}(V^\dagger)
\ee
whose localization to the half-BPS sector yields the expression \eqref{heavydeterminant} for $U = e^{M_1}$ and $V^\dag = e^{-M_2}$. While the identity \eqref{deltaexpansionchar} does not descend directly to a delta function for Hermitian matrices, the delta function-like properties of \eqref{heavydeterminant} can be seen via its effect on the saddle point equations of the half-BPS matrix models. Diagonalising the matrices $M_{1}$ and $M_2$ in terms of the eigenvalues $\{ x_i\}_{i=1}^N$ and $\{ y_i\}_{i=1}^N$, respectively, we have
\be \label{eq: heavydeterminant eigen}
\sum_R \tr_R(e^{M_1}) \tr_R(e^{-M_2}) = \prod_{i,j=1}^N \frac{1}{1 - e^{x_i - y_j}}.
\ee
This operator exhibits poles whenever the two eigenvalues $x_i$ and $y_j$ collide, so the coupled matrix model action is minimized for configurations in which pairs of $x_i$ and $y_j$ eigenvalues are identified modulo permutations.

In our analysis, however, we find that the operator \eqref{heavydeterminant} alone does not lead to the appropriate saddle point in the matrix model. A natural modification that will be relevant for us is a ``supersymmetric'' extension involving both bosonic and fermionic determinants:
\ie\label{susicdelta}
\widehat{\delta}_{12} &= \left( \sum_R \Tr_R (e^{M_1}) \Tr_R (e^{-M_2}) \right) \left( \sum_{R} \Tr_{R^T} (e^{M_1}) \Tr_{R} (e^{-M_2}) \right) \\
&= \frac{\det \left(\mathds{1} \otimes \mathds{1} + e^{M_1} \otimes e^{-M_2}  \right)}{\det \left(\mathds{1} \otimes \mathds{1} - e^{M_1} \otimes e^{-M_2}  \right)}
\fe
where $R^T$ denotes the transpose of Young diagram labelling the irrep $R$. This operator can also be expanded in terms of Hook-Schur polynomials $HS_R$ which are characters of the irreps of the supergroup $U(N|N)$:\footnote{For an even number of such operators, the formula appears more symmetric, see Eq. 2 of~\cite{berele1985hook}.}
\be
\widehat{\delta}_{12} = \sum_R HS_R(e^{M_1}| 0) HS_R(e^{-M_2}|e^{-M_2})
\ee
In the following, we refer to $\widehat{\delta}_{12}$ as the \textit{delta operator}.

\begin{figure}[t]
\centering
\includegraphics[width=80mm]{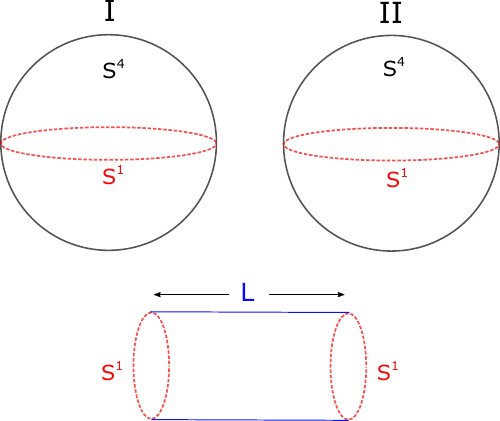}
\caption{The conformal boundary of the bubbling wormhole is given by the zero-length $\ell \to 0$ limit of the above ``plumbed'' geometry. The result is two four-spheres identified along a common $S^1$ on which the delta operator is placed.}
\label{fig:loops}
\end{figure}

The delta operator $\widehat{\delta}_{12}$ admits a natural interpretation from the perspective of a $U(N)_1 \times U(N)_2$ gauge theory (Figure \ref{fig:loops}). Consider integrating out a pair of massive bifundamental fields: a boson and a fermion both in the $(N,\bar{N})$ representation and of mass $\ell$. Upon localization, integrating out the boson produces an inverse determinant $\det^{-1} (\mathds{1} \otimes \mathds{1} - e^{-\ell} \, e^{M_1} \otimes e^{-M_2} )$ while integrating out the fermion produces a determinant $\det (\mathds{1} \otimes \mathds{1} + e^{-\ell} \, e^{M_1} \otimes e^{-M_2} )$ in the numerator  \cite{Gomis:2006sb,Gomis:2006im}. The ratio of determinants can be written as a plethystic exponential of double-trace operators:
\be
\exp \left( \sum_{n=1}^\infty \frac{e^{-n \ell}}{n} \left( 1 + (-1)^{n+1} \right) \Tr (e^{n M_1}) \Tr (e^{-n M_2}) \right),
\ee
where the $\Tr$ in this expression is in the fundamental representation. The delta operator $\widehat{\delta}_{12}$ arises in the $\ell \to 0$ limit, which suggests a possible UV origin for our construction.\footnote{Our construction bears similarity with an example in topological string theory \cite{Aganagic:2002qg}. There, a system of several $U(N_i)$ Chern-Simons theories on separate $S^3$ is connected in a cyclic chain via cylinder amplitudes, and operators that link the $S^3$'s along $S^1$'s are sums over representations of cyclic pairs of Wilson lines.}

We now describe the bubbling wormhole (BW) geometries. The half-BPS Wilson loop preserves a $SO(2,1) \times SO(3) \times SO(5)$ subgroup of the $SU(2,2|4)$ symmetry of $\cN = 4$ SYM, and the dual supergravity solutions respect these isometries via the metric ansatz
\be
ds^2 = f_1^2 ds_{\AdS_2}^2 + f_2^2 ds_{S^2}^2 + f_4^2 ds_{S^4}^2 + ds_{\Sigma}^2,
\ee
where the metric functions depend on the coordinate $z$ of a Riemann surface $\Sigma$ \cite{DHoker:2007mci}. The solutions are determined by a pair of harmonic functions $h_1$ and $h_2$ on $\Sigma$, subject to regularity conditions. For a single asymptotically $\AdS_5 \times S^5$ region, $\Sigma$ corresponds to the lower half-plane with a pole at infinity.

The bubbling wormhole geometries we consider have multiple poles on $\pa\Sigma$, which correspond to multiple asymptotically $\AdS_5 \times S^5$ regions. These geometries turn out to be multi-covers of $\AdS_5 \times S^5$; they are locally $\AdS_5 \times S^5$ but differ globally. The conformal boundary consists of a union of multiple four-spheres $\cup_A S_A^4$ that intersect on a common $\cap_A S_A^4 = S^1$, which is the boundary of the ${\rm EAdS}_2$ fiber present at each point in $\Sigma$. Notably, the conformal boundary is connected, so these are \textit{not} Euclidean wormholes in the sense of having completely disconnected boundaries.

A key feature of these geometries is the presence of a codimension-2 conical singularities in the interior of $\Sigma$, located at the fixed points of the covering map. These singularities have the local form $\AdS_2 \times S^2 \times S^4$ and correspond to a conical excess of $2 \pi$ (which is a total angle $4 \pi$ for the two-cover). Such singularities require a negative-energy source for the geometry to satisfy the supergravity equations of motion. We model the source as a codimension-2 cosmic brane of negative tension on $\AdS_2 \times S^2 \times S^4$. This is consistent with the attractive nature of the delta operator $\widehat\delta_{12}$ whose free energy $F_\delta \sim -N^2/\sqrt\lambda$ is negative. On the bulk side, the leading contributions $\sim N^2$ from the cosmic brane tension and the conical singularity in the Einstein-Hilbert action cancel. The subleading $\sim N^2/\sqrt\lambda$ behavior can be accommodated by a Dvali-Gabadadze-Porrati term \cite{Dvali:2000hr}, i.e. induced gravity on the bulk source, in the worldvolume model, though the precise coefficient remains undetermined.

Conical excess geometries with opening angle $2\pi\alpha$, $\alpha > 1$, are typically associated with negative-tension sources and do not admit an ordinary orbifold CFT description. In the special case $\alpha \in \mathbb{Z}_{>1}$, a controlled string-theoretic realization can be obtained by combining a discrete identification with orientifold planes, whose negative tension and charge support the excess while preserving consistency through tadpole cancellation.

In our setup the excess is $2\pi$ at the fixed point of the two-cover, localized on $\AdS_2 \times S^2 \times S^4$ inside $\AdS_5 \times S^5$. Two features constrain a direct orientifold interpretation: First, our solutions have constant axio-dilaton, which among type IIB orientifolds singles out the O3-plane as the natural candidate. Second, an O3-plane is codimension-six, while our singular locus is codimension-two and wraps the full $S^4$. An O3 source would therefore have to be smeared along the $S^4$ directions. Smeared orientifolds have been previously considered as effective descriptions in flux compactifications, but whether a smeared O3 distribution on $S^4$ admits a fully localized completion compatible with the supergravity equations and charge quantization is less clear. Identifying the microscopic source, whether smeared O3-planes, a different brane configuration, or an effective DGP-type description without a string-theoretic origin, remains an open question.

The paper is organized as follows. In Section \ref{section: bubbling wormholes}, we review the construction of bubbling geometries and present the two-cover and four-cover bubbling wormhole solutions. Their properties are determined from the harmonic functions given in eqns.~\eqref{twoboundarysolution} and~\eqref{fourboundarysolutions}. In Section~\ref{WilsonLoopreview}, we review the properties of the matrix models that describe $1/2$ BPS Wilson loops in $\mathcal{N}=4$ SYM after performing the supersymmetric localization, and in particular, how the matrix model resolvent can be mapped to the harmonic functions that describe the dual bubbling geometry. In Section \ref{MMbubbling}, we analyze the matrix models whose spectral curves match the supergravity harmonic functions. In particular we notice that the resolvents of a certain class of Gaussian–Penner matrix models \cite{Tan:1991ay}, discussed in more detail in Appendix~\ref{appendix: Gaussian-Penner model}, lead precicely to the harmonic functions describing the two-cover and four-cover bubbling wormholes. Our perspective is that the Gaussian–Penner matrix models constitute an effective matrix model description of such geometries, with the reason being that they do not clarify the rules by which one can compute observables on top of the large-$N$ saddle (for example, one can further insert probe Wilson loops on top of the large-$N$ backreacted saddles, see Section~\ref{sec: probe loops}). By contrast, the two-matrix model \eqref{2-cover WH matrix model eigenvalue basis}, together with the coordinate transform \eqref{1-2 coordinate transform}, constitutes our fundamental proposal for how the two-cover bubbling wormhole emerges from two originally decoupled Gaussian matrix models after the insertion of the supersymmetric delta operator \eqref{susicdelta}. In Section~\ref{fourcovercyclicdeltas} we generalise this construction to the four-cover case. In Section \ref{sec: free energy}, we compute the free energy of the delta operator and discuss its bulk interpretation, specifically the properties of the resulting negative energy source. In Section \ref{sec: probe loops}, we analyze probe fundamental strings and Wilson loop operators and show how the results from the bulk computations are in agreement with those from the dual (multi) matrix model (including the delta-operator insertions). Further details on the Gaussian-Penner effective matrix models whose spectral curve coincides with the supergravity harmonic functions are collected in Appendix \ref{appendix: Gaussian-Penner model}.

\section{Bubbling wormholes} \label{section: bubbling wormholes}

\subsection{Geometries dual to half-BPS Wilson loops} \label{preliminarybubbling}

We now review the construction of bubbling geometries dual to half-BPS Wilson loops in irreducible representations, developed in \cite{DHoker:2007mci} (see also \cite{Yamaguchi:2006te} and \cite{Okuda:2008px,Benichou:2011aa,Aguilera-Damia:2017znn} for reviews).

The work \cite{DHoker:2007mci} reduced the problem of finding supergravity solutions dual to half-BPS Wilson loops to that of finding a pair of harmonic functions, $h_1$ and $h_2$, on the lower-half-plane or a disk $\Sigma$, that satisfy certain regularity conditions.

A half-BPS Wilson loop in $\mathcal{N}=4$ super Yang-Mills preserves a $SO(2,1) \times SO(3) \times SO(5)$ subgroup of the bosonic part of the superconformal group $SO(2,4) \times SO(6)$. Therefore, bubbling geometry backgrounds in dual IIB supergravity preserve a $SO(2,1) \times SO(3) \times SO(5)$ isometry group. Geometries with this isometry group are given by the metric ansatz\footnote{The solutions we present can be in either Lorentzian or Euclidean signature, but we primarily discuss the Euclidean case with conformal boundary $S^4$ to make direct contact with the matrix model. This only changes signs in the $\AdS_2$ component of the metric and makes $\AdS_5$ components of $F_5$ pure imaginary.}
\begin{equation} \label{eq: metric ansatz}
    ds^2 = f_1^2 ds_{\AdS_2}^2 + f_2^2 ds_{S^2}^2 + f_4^2 ds_{S^4}^2 + ds_{\Sigma}^2,
\end{equation}
where the $\AdS_2$, $S^2$, and $S^4$ components of the solution are nontrivially fibered over the disk $\Sigma$. We write the metric on $\Sigma$ using complex coordinates $(z,\bar{z})$ as
\be
ds_{\Sigma}^2 = 4 \rho^2 dz d\bar{z}.
\ee
We take $\Sigma$ to be the lower half $z$-plane with boundary $\partial\Sigma$ along the real line (including infinity). There are many valid choices of local coordinates, but these coordinates simplify the relevant functions for bubbling wormholes.

We first describe in words the geometry and fluxes of a bubbling solution dual to half-BPS Wilson loops. Such solutions are completely determined by a pair of harmonic functions $h_{1}$ and $h_{2}$ on $\Sigma$. The zero loci of $h_2$ define the boundary $\partial\Sigma$ of $\Sigma$. On the $z$-plane, the zero loci of $h_2$ and thus $\partial\Sigma$ lie along the real line. $h_2$ does not have branch cuts. Simple poles of (the holomorphic parts of) $h_1$ and $h_2$ on $\partial\Sigma$ correspond to asymptotically $\AdS_5 \times S^5$ regions.

Away from such singularities on $\partial\Sigma$, the sizes of $S^2$ or $S^4$ vanish in an alternating fashion in a way that is dependent on the structure of the branch cuts of $h_1$. In particular, the metric components of $S^2$ and $S^4$ respectively vanish on and off the branch cuts. This indicates that $\partial\Sigma$ is not a true boundary of the full ten-dimensional solution.

The fact that $S^2$ and $S^4$ sizes shrink on $\partial\Sigma$ indicates that there are nontrivial $S^3$ and $S^5$ cycles in the bubbling geometry. A $S^3$ cycle is a contour on $\Sigma$ that starts and ends on different branch cuts on $\partial\Sigma$. Similiarly, a $S^5$ cycle is a contour on $\Sigma$ that starts and ends on regions in between branch cuts on $\partial\Sigma$. There are various NS-NS and R-R fluxes across the $S^3$ and $S^5$ cycles, reflecting the presence of D5 and D3 branes dissolved into the solution. We can also have F1 charges by taking the contours to instead represent 7-cycles $S^3 \times S^4$ or $S^2 \times S^5$.

\begin{figure}[t]
\centering
\includegraphics[width=140mm]{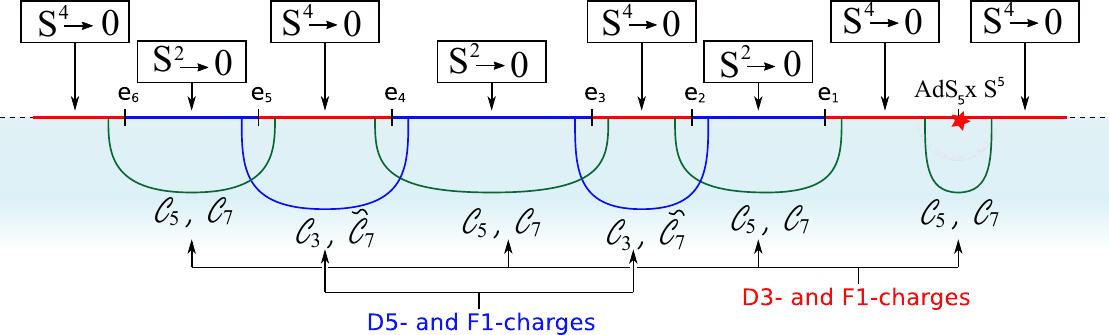}
\caption{The bulk geometry is dictated by the Riemann surface $\Sigma$ and the number of cuts (blue) and poles (star) the functions $h_1$ and $h_2$ has on $\partial\Sigma$. Here we depict the usual case of a single pole (i.e. one asymptotically $\AdS_5 \times S^5$ region) and multiple cuts, which is a geometry dual to a single backreacted Wilson loop in a large representation $R$ with order $N^2$ number of boxes. $\mathcal{C}_D$ denote nontrivial cycles in the geometry, where $D$ is the dimension of the cycle.}
\label{fig:SigmaCyles}
\end{figure}

The metrics and fluxes are written in terms of the harmonic functions $h_{1}$ and $h_{2}$ \cite{DHoker:2007mci}. It will sometimes be convenient to write $h_{1,2}$ in terms of holomorphic functions $\mathcal{A}$, $\mathcal{B}$ on $\Sigma$ such that
\be\label{holomorphicmetricfunctions}
h_1 = \mathcal{A} + \bar{\mathcal{A}}, \qquad h_2 = \mathcal{B} + \bar{\mathcal{B}}.
\ee
The metric functions $f_1$, $f_2$, $f_4$, and $\rho$, as well as the dilaton\footnote{The closed string dilaton is $\Phi = 2 \phi$.}, are given by
\begin{align}\label{AnsatzfunctionsI}
    f_1^4 &= -4 e^{+2 \phi} h_1^4 \frac{W}{N_1} \nonumber \\
    f_2^4 &= +4 e^{-2 \phi} h_2^4 \frac{W}{N_2} \nonumber \\
    f_4^4 &= +4 e^{-2 \phi} \frac{N_2}{W} \nonumber \\
    \rho^8 &= - \frac{W^2 N_1 N_2}{h_1^4 h_2^4} \nonumber \\
    e^{4 \phi} &= -\frac{N_2}{N_1} \, ,
\end{align}
where
\begin{align}\label{AnsatzfunctionsII}
    W &= \partial_{z} h_1 \partial_{\bar{z}} h_2 + \partial_{z} h_2 \partial_{\bar{z}} h_1 \nonumber \\
    V &=\partial_{z} h_1 \partial_{\bar{z}} h_2 - \partial_{z} h_2 \partial_{\bar{z}} h_1 \nonumber \\
    N_1 &= 2 h_1 h_2 \partial_z h_1 \partial_{\bar{z}} h_1 - h_1^2 W \nonumber \\
    N_2 &= 2 h_1 h_2 \partial_z h_2 \partial_{\bar{z}} h_2 - h_2^2 W \, .
\end{align}

The fluxes require a few more definitions that are found in \cite{Benichou:2011aa,Aguilera-Damia:2017znn}. The relevant NS-NS and R-R fluxes are
\begin{align}
    H_3 &= dB_2 \nonumber \\
    F_3 &= dC_2 \nonumber \\
    F_5 &= dC_4 + \frac{1}{8} \left( B_2 \wedge F_3 - C_2 \wedge H_3 \right),
\end{align}
with potentials
\begin{align}
    B_2 &= b_1 \hat{e}_{\AdS_2} \nonumber \\
    C_2 &= b_2 \hat{e}_{S^2} \nonumber \\
    C_4 &= - 4 j_1 \hat{e}_{\AdS_2} \wedge \hat{e}_{S^2} + 4 j_2 \hat{e}_{S^4},
\end{align}
where $\hat{e}$ are unit volume elements of the corresponding spaces. The potentials are defined in terms of the functions
\begin{align}
    b_1 &= -2 i \frac{h_1^2 h_2 V}{N_1} - 2 \tilde{h}_2 \nonumber \\
    b_2 &= -2 i \frac{h_1 h_2^2 V}{N_2} + 2 \tilde{h}_1 \nonumber \\
    j_2 &= i h_1 h_2 \frac{V}{W} - \frac{3}{2} \left( \tilde{h}_1 h_2 - h_1 \tilde{h}_2 \right) + 3 i \left( \mathcal{C} - \bar{\mathcal{C}} \right) \nonumber \\
    \partial j_1 &= -i \frac{f_1^2 f_2^2}{f_4^4} \partial j_2 + \frac{1}{8} \left( b_1 \partial b_2 - b_2 \partial b_1 \right),
\end{align}
where $\tilde{h}_1 = i(\mathcal{A} - \bar{\mathcal{A}})$, $\tilde{h}_2 = i(\mathcal{B} - \bar{\mathcal{B}})$, and $d\mathcal{C} = \mathcal{B} \partial \mathcal{A} - \mathcal{A} \partial \mathcal{B}$.

There are scaling transformations for the harmonic functions $h_{1,2}$ that rescale the metric, dilaton, and fluxes. One transformation is $h_{1,2} \to c^2 h_{1,2}$, which keeps the dilaton $\phi$ invariant and takes
\ieg
    \rho^2 \to c^2 \rho^2, \qquad f_{1}^2 \to c^2 f_{1}^2, \qquad f_{2}^2 \to c^2 f_{2}^2, \qquad f_{4}^2 \to c^2 f_{4}^2 \\
    H_3 \to c^{-1} H_3, \qquad F_3 \to c^{-1} F_3, \qquad F_5 \to c^{-1} F_5\, .
\feg
Another scaling transformation is $h_{1} \to c^{-1} h_{1}$ and $h_{2} \to c h_{2}$, which leaves metrics and fluxes invariant and takes
\be\label{dilaton scaling}
e^{2 \phi} \to c^2 e^{2 \phi}.
\ee
All supergravity solutions considered in this work have a constant dilaton $\phi=\phi_0$. We will use \eqref{dilaton scaling} to normalize $h_{1,2}$, so that $g_s=e^{2\phi}=1$.

Further conditions on $h_{1,2}$ that are required for regularity of the supergravity solution are as follows:
\begin{itemize}
\item $h_1$ may vanish only on the segments of $\partial\Sigma$ where $S^4$ vanish (i.e. off the branch cut). Segments of $\partial\Sigma$ where $h_1 \neq 0$ correspond to the vanishing of $S^2$.
\item $h_1$ and $h_2$ are positive definite in the interior of $\Sigma$, but can vanish on $\partial\Sigma$.
\item The differentials $\partial_z h_{1}$ and $\partial_z h_{2}$ share common zeroes in the interior of $\Sigma$, when such zeroes exist.
\end{itemize}
The final regularity condition was not required for bubbling solutions with a single $\AdS_5 \times S^5$ region, as $\partial_z h_{1,2}$ do not have zeroes in the interior in that case. If we change the boundary conditions so that there are multiple poles on $\pa \Sigma$ corresponding to $\AdS_5 \times S^5$ regions, the differentials $\partial_z h_{1,2}$ acquire zeroes in the interior and the zeroes need to coincide for the regularity of the solution. This regularity condition was discussed in \cite{DHoker:2007hhe,DHoker:2007zhm} in the context of bubbling Janus solutions.

In our work, we consider a more general class of boundary conditions for $h_1$ and $h_2$ that involve multiple poles on $\pa \Sigma$ corresponding to multiple asymptotically-$\AdS_5 \times S^5$ regions, whose conformal boundaries $S_i^4$ intersect on a common $S^1$: $\cap_i \, S_i^4 = S^1$.

The resulting wormhole-like geometries we consider turn out to have vanishing $H_3$ and $F_3$ fluxes and constant dilaton $\Phi = 2\phi$. In the context of bubbling geometries dual to half-BPS Wilson loops, it can be shown that a constant dilaton actually implies a locally $\AdS_5 \times S^5$ solution \cite{DHoker:2007mci}. We find that our solutions are multiple covers of the usual $\AdS_5 \times S^5$ that differ only globally from $\AdS_5 \times S^5$. The conformal boundaries $S_i^4$ of multiple asymptotically-$\AdS_5 \times S^5$ regions intersect on a common $S^1$: $\cap_i \, S_i^4 = S^1$. Due to being a multi-cover, these geometries possess a finite number of loci of codimension-two conical singularities of the form $\AdS_2 \times S^2 \times S^4$ that are located in the interior of $\Sigma$ but that stretch out to the common $S^1$ of its conformal boundary $\cup_i S_i^4$.

\subsection{Review of $\AdS_5 \times S^5$ solution}
\label{vacuum AdS solution}

Let us review the construction of the $\AdS_5 \times S^5$ solution in the bubbling ansatz preserving the symmetries of the half-BPS Wilson loop. The vacuum solution corresponds to the insertion of a loop in the trivial representation, i.e. no loop, so the ansatz simply amounts to an $\AdS_2 \times S^2 \times S^4$ slicing of $\AdS_5 \times S^5$.

\begin{figure}
    \centering
    \includegraphics[width=0.7\linewidth]{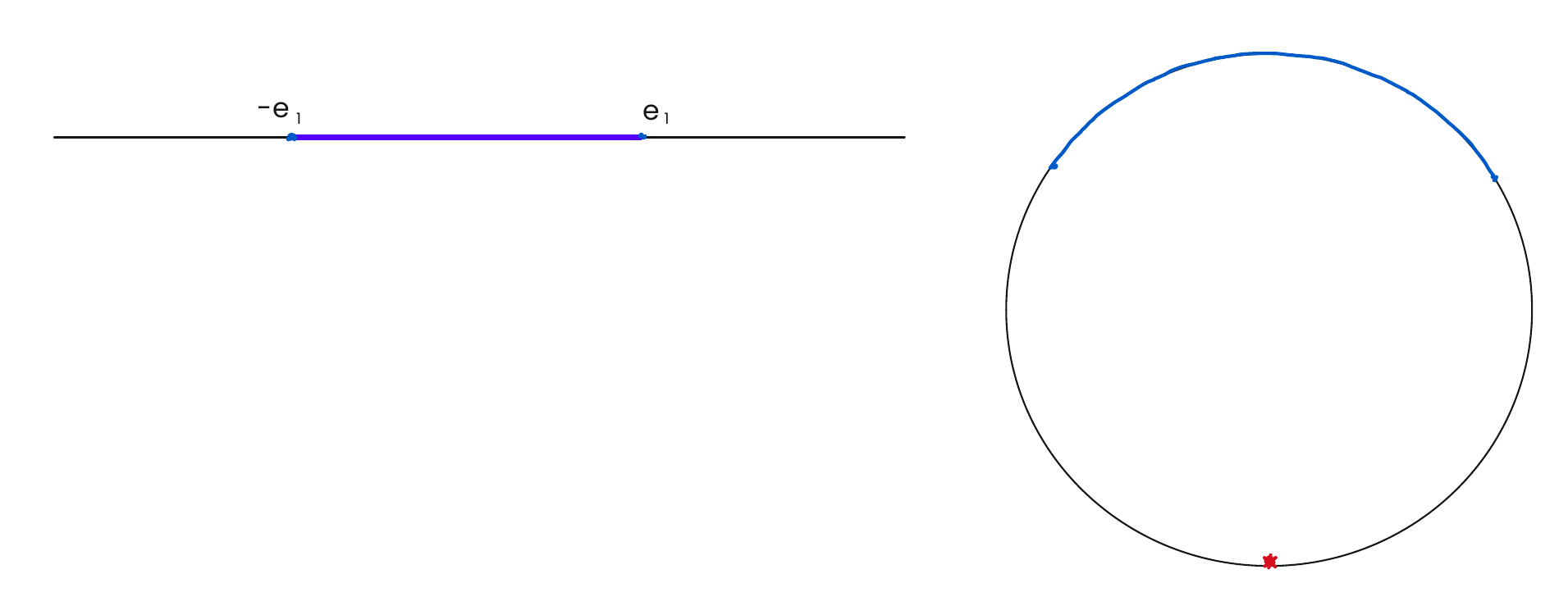}
    \caption{The Riemann surface $\Sigma$ describing $\AdS_5 \times S^5$ can be viewed as the lower half-plane with a single branch cut and a simple pole at infinity, which can be mapped to a disk containing one cut and one simple pole.}
    \label{fig:AdS5disk}
\end{figure}

The harmonic functions for $\AdS_5 \times S^5$ are
\begin{align}\label{oneboundarysolution}
    h_1 &= \frac{\alpha'}{4} \sqrt{b^2 - z^2} \ + \ \text{c.c.} \nonumber \\
    h_2 &= i \frac{\alpha'}{4} z  + \ \text{c.c.}
\end{align}
where $b$ is a constant whose value will be determined in terms of the physical parameters. The pole corresponding to the asymptotic region is at $z = - i \infty$. The lower-half $z$-plane, consisting of a single branch cut and a simple pole, and its mapping to a disk with one cut and one pole are depicted in Figure~\ref{fig:AdS5disk}.

To show that these harmonic functions give a slicing of $\AdS_5 \times S^5$, it is convenient to work on the $w$-plane where $z = -i b \sinh{w}$ and we take $w = x + i y$. One finds the $\AdS_2 \times S^2 \times S^4$ slicing of $\AdS_5 \times S^5$
\begin{equation} \label{eq: ads2 x s2 x s4 slicing of pure AdS}
    ds_{\AdS_5 \times S^5}^2 = L^2 \bigg[ \cosh^2{x} \ ds_{\AdS_2}^2 + \sinh^2{x} \ ds_{S^2}^2 + \cos^2{y} \ ds_{S^4}^2 + (dx^2 + dy^2) \bigg],
\end{equation}
where $L^2 = \alpha' b$ and $\Sigma$ becomes a semi-infinite strip $x \geq 0$ and $y \in \left[ -\frac{\pi}{2}, \frac{\pi}{2} \right]$. The prefactor of $ds_{S^4}^2$ differs from the usual one due to an unusual choice for the range of $y$.

One can observe from \eqref{eq: ads2 x s2 x s4 slicing of pure AdS} that it is possible to reach the $S^1$ part of the conformal boundary anywhere inside $\Sigma$ due to always having an $\AdS_2$ fiber at each point of $\Sigma$. However, the full conformal boundary $S^4$ is accessible only as $x \to +\infty$ or $z \to - i \infty$. The bubbling wormhole geometries that we present shortly will also have the property that one can reach a common $S^1$ of the linked $S_i^4$'s via the $\AdS_2$ fiber. In general, there can be several asymptotic $\AdS_5 \times S^5$ poles along $\partial\Sigma$ whose conformal boundaries are distinct $S^4$'s.

Let us directly compute the five-form flux across $S^5$. We work in the Einstein frame where $L^4 = 4 \pi \alpha'^2 N$. The D3 brane charges at asymptotic regions are given by
\be
Q_{D3} = \int_{S^5} dC_4 = \widehat{\text{Vol}}(S^4) \int_{\gamma} 4 \partial j_2,
\ee
where $\gamma$ is a counter-clockwise contour on $\Sigma$ centered at the origin with radius greater than $b$ that starts and ends on $\partial\Sigma$. $\widehat{\text{Vol}}(S^4) = \frac{8 \pi^2}{3}$ is the unit volume of $S^4$. The D3 charges are related to the number of D3 branes as $Q_{D3} = N (4 \pi^2 \alpha')^2$. A direct calculation of $Q_{D3}$ using the definitions in Section \ref{preliminarybubbling} yields
\begin{equation} \label{eq: charge for one boundary}
    Q_{D3} = \frac{3 \pi}{2} \alpha'^2 b^2 \, \widehat{\text{Vol}}(S^4)
\end{equation}
Then we have
\be
N = \frac{b^2}{4\pi}
\ee
and $L^4 = \alpha'^2 b^2$. The units of five-form flux grows quadratically with the size of the cut, and the expressions for $N$ and $L^4$ indicate that $b = \sqrt{\lambda}$.

\subsection{Two-cover geometry}
\label{tworegions}

In this section, we consider geometries with two points on $\partial\Sigma$ where the metric asymptotes to $\AdS_5 \times S^5$. We refer to this geometry as the two-cover bubbling wormhole (BW$_2$) because it is a double cover\footnote{It is interesting to note that a somewhat different double-cover structure appears in bubbling geometries for $\AdS_2\times S^2$ \cite{Lunin:2015hma}, in certain constructions describing global $\AdS_2$ with two disconnected boundaries. In our work $\EAdS_2$ has the topology of a disk with a single boundary and our double cover structure is of a different nature.} of global $\AdS_5 \times S^5$ with conformal boundary a pair of $S_i^4$'s ($i=1,2$) that are linked along a common $S^1$.

We work on the lower-half $z$-plane $\Sigma$ and impose that $\partial\Sigma$ lies on the real axis. Solutions with two poles are quite constrained due to the regularity conditions on $h_{1,2}$. In particular, the condition that the differentials $\partial_z h_{1}$ and $\partial_z h_{2}$ share common zeroes in the interior of $\Sigma$ rules out any one-cut configuration with two simple poles for $h_{1,2}$.

\begin{figure}
    \centering
    \includegraphics[width=0.7\linewidth]{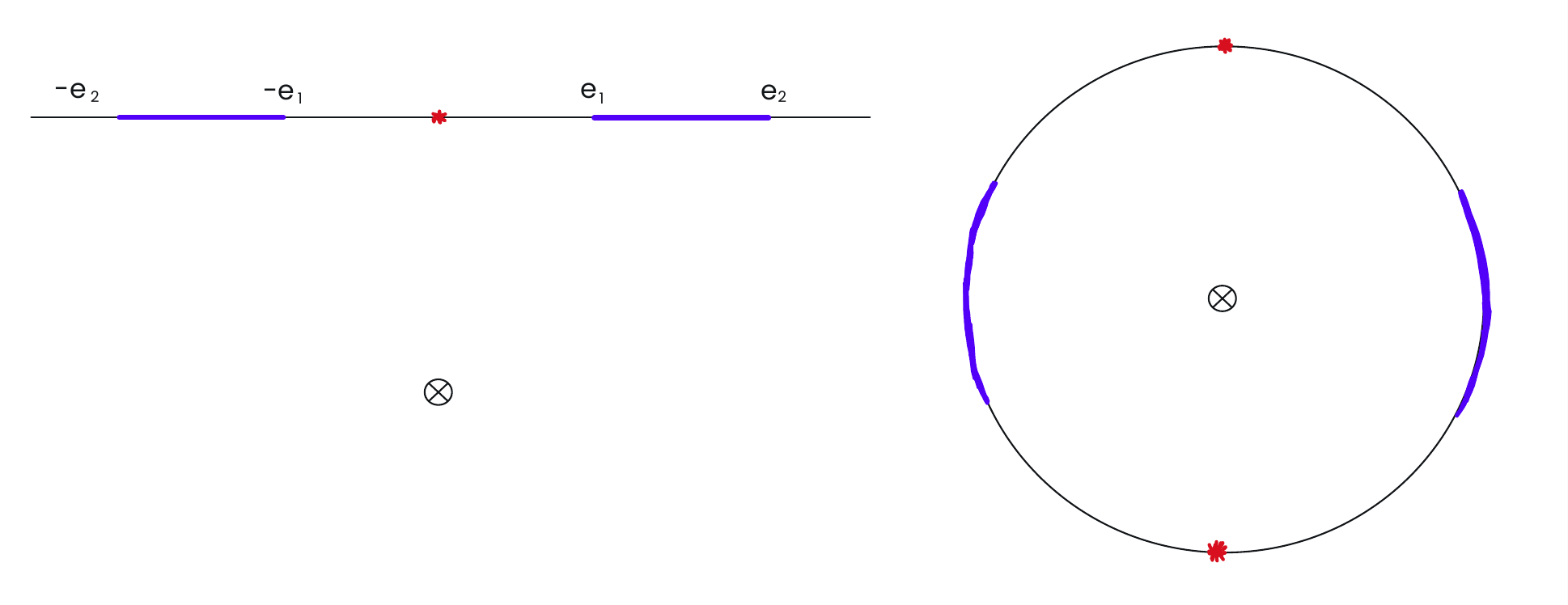}
    \caption{The Riemann surface $\Sigma$ describing BW$_2$ can be viewed as the lower half-plane with two branch cuts and two simple poles, which can be mapped to a disk containing two cuts and two simple poles. The symbol $\otimes$ denotes the conical singularity (excess) on $\Sigma$.}
    \label{fig:DoubleCover}
\end{figure}

We find a class of two-cut solutions that satisfy the regularity conditions. In these solutions, the cuts are placed in a symmetric manner on each side of the simple poles. If we fix the asymptotic $\AdS_5 \times S^5$ points to be located at $z=0$ and $z=\infty$, the harmonic functions satisfying the regularity conditions are given by
\begin{align}\label{twoboundarysolution}
    h_1 &= \frac{\alpha'}{4} \sqrt{\frac{(e_2^2 - z^2)(z^2 - e_1^2)}{z^2}} \ + \ \text{c.c.} \nonumber \\
    h_2 &= i \frac{\alpha'}{4} \left( z - \frac{e_1 e_2}{z} \right) \ + \ \text{c.c.},
\end{align}
where $0 < e_1 < e_2$. The holomorphic part $\cA(z)$ of $h_1$ has branch cuts along $z=[-e_2,-e_1] \cup [e_1,e_2]$. The lower-half $z$-plane, consisting of two branch cuts and two simple poles, and its mapping to a disk with two cuts and two poles are depicted in Figure~\ref{fig:DoubleCover}.

The explicit expressions of the metric functions $f_1^2$, $f_2^2$, $f_4^2$, and $\rho$, as well as the dilaton $\phi$ and $b_1$, $b_2$, are readily computed from the formulas in Section \ref{preliminarybubbling}. It can be found that the dilaton is constant and that $b_1 = b_2 = 0$, leading to the presence of only five-form flux $F_5$ on the background.

Though the exact expressions for the non-trivial metric functions are rather involved, we can study their asymptotic behaviors near the $\AdS_5 \times S^5$ regions. We now take $z = -i e^w$ and $w = x + i y$. The two-cover wormhole gets mapped to the infinite strip $-\infty < x < \infty$ and $y \in \left[ -\frac{\pi}{2}, \frac{\pi}{2} \right]$. The asymptotic expressions for $x \to + \infty$ are
\begin{align}
    f_1^2 \sim f_2^2 &\sim \alpha' \frac{e_1^2 e_2^2}{e_2 - e_1} e^{2 x} \nonumber \\
    f_4^2 &\sim \alpha' (e_2 - e_1) \cos^2{y} \nonumber \\
    4 \rho^2 &\sim \alpha' (e_2 - e_1),
\end{align}
while those for $x \to - \infty$ are
\begin{align}
    f_1^2 \sim f_2^2 &\sim \alpha' \frac{1}{e_2 - e_1} e^{-2 x} \nonumber \\
    f_4^2 &\sim \alpha' (e_2 - e_1) \cos^2{y} \nonumber \\
    4 \rho^2 &\sim \alpha' (e_2 - e_1).
\end{align}
To bring this into famililar form, let us define new radial coordinates $x_+ = x - \log( \frac{e_2 - e_1}{2} )$ and $x_- = x + \log ( \frac{e_2 - e_1}{2 e_1 e_2} )$ in the respective asymptotic regions. Plugging into the metric ansatz \eqref{eq: metric ansatz}, the metric acquires the following form in the two asymptotic regions 
\be
ds_{\pm}^2 \sim \alpha' (e_2 - e_1) \bigg[ \frac{1}{4} e^{2 |x_{\pm}|} \left( ds_{\AdS_2}^2 + ds_{S^2}^2 \right) + dx_{\pm}^2 + dy^2 + \cos^2{y} \ ds_{S^4}^2 \bigg].
\ee
These expressions for the asymptotic metric agree with the asymptotic form of \eqref{eq: ads2 x s2 x s4 slicing of pure AdS} if we make the identification
\be
L_i^4 = \alpha'^2 (e_2 - e_1)^2
\ee
where $L_i$ is the $\AdS$ radius in each asymptotically-$\adsS$ region. We find that the sizes of the two asymptotic $\AdS_5 \times S^5$ regions are equal and depend only on the difference of the locations of the two branch points of $h_1$.

Let us check the above conclusion via a direct calculation of the five form flux across $S^5$ in the Einstein frame where $L^4 = 4 \pi \alpha'^2 N$. The D3 brane charges at asymptotic regions are given by
\be
Q_{D3}^{(i)} = \int_{S_i^5} dC_4 = \widehat{\text{Vol}}(S^4) \int_{\gamma_i} 4 \partial j_2,
\ee
where $\gamma_i$ is now a sufficiently small counter-clockwise contour on $\Sigma$ that ends on $\partial\Sigma$ and surrounds the $i$-th asymptotic $\AdS_5 \times S^5$ pole. Sufficiently small means that the contour does not touch other cuts or enclose other poles. Now the D3 charges are related to the number of D3 branes by $Q_{D3}^{(i)} = N^{(i)} (4 \pi^2 \alpha')^2$. A direct calculation of $Q_{D3}^{(i)}$ yields
\begin{equation} \label{eq: charge for two boundaries}
    Q_{D3}^{(i)} = \frac{3 \pi}{2} \alpha'^2 (e_2 - e_1)^2 \widehat{\text{Vol}}(S^4)
\end{equation}
for both asymptotic regions of the two-cover wormhole. Namely, the signs of charges computed around both poles are the same. The D3 charges computed from a counter-clockwise contour surrounding each branch cut of $h_1$ on $\partial\Sigma$ gives minus the charge \eqref{eq: charge for two boundaries}. We then find
\be
N^{(i)} = \frac{1}{4\pi} (e_2 - e_1)^2
\ee
and
\be
L_i^4 = \alpha'^2 (e_2 - e_1)^2.
\ee
Hence, a direct computation of the flux across the $S^5$ agrees with the analysis from the asymptotic metric. We find that the 't Hooft coupling $\lambda = 4 \pi g_s N = L^4/\alpha'^2$ is
\be\label{2-cover WH lambda = e2 - e1}
\lambda = (e_2 - e_1)^2,
\ee
and that the branch points individually scale as $e_1 \sim e_2 \sim \sqrt{\lambda}$.

We now describe some properties of the two-cover wormhole. The background value of the dilaton is constant $g_s^2 = e^{4 \phi} = 1$ in our normalization of $h_{1,2}$. This means that we are working in the normalization where the ten-dimensional Newton's constant in the Einstein and string frames are the same. It can be shown \cite{DHoker:2007mci} for these half-BPS solutions that constant $\phi$ implies that the solution is locally $\AdS_5 \times S^5$ for bubbling geometries dual to half-BPS Wilson loops, and indeed our solution has the global structure of its double cover.

While there are two distinct asymptotically-$\adsS$ regions on $\Sigma$, the conformal boundaries $S_i^4$ associated to the $(\adsS)_i$ regions intersect on a common $S^1$. In an $\EAdS_2 \times S^2$ slicing of Euclidean $\AdS_5$, the $\EAdS_2$ slices share a common conformal boundary $S^1 \subset S^4$. The $\AdS_2$ fibers present at every point on $\Sigma$ of the 2-cover wormhole inherit the property that they share a common $S^1$ boundary. Therefore, the conformal boundary of the 2-cover wormhole is
\be
S_1^4 \cup S_2^4, \quad {\rm where} \quad S_1^4 \cap S_2^4 = S^1.
\ee
There is only one connected conformal boundary, and there is no obstruction to traversing from one asymptotically-$\adsS$ region to another via $\Sigma$ in the bulk.

Importantly, this geometry has a codimension-two locus of conical singularities of the form $\AdS_2 \times S^2 \times S^4$ in the interior of $\Sigma$. The singularity is located at
\be
z_c = -i \sqrt{e_1 e_2}
\ee
and corresponds to the fixed point of the $\mathbb{Z}_2$ isometry $z \to -\frac{e_1 e_2}{z}$. Let us expand the metric on $\Sigma$
\be
\frac{1}{\alpha'} ds_\Sigma^2 = \frac{ (e_2 - e_1) (z^2 + e_1 e_2) (\bar{z}^2 + e_1 e_2)}{z \bar{z} \sqrt{(e_2^2 - z^2)(z^2 - e_1^2)(e_2^2- \bar{z}^2)(\bar{z}^2 - e_1^2)}} dz d\bar{z}
\ee
near $z_c$. With $z = z_c + r e^{i \theta}$, we find near $r=0$ that
\be
\frac{1}{\alpha'} \left.ds_\Sigma^2\right\rvert_{z_c} \approx \frac{4 (e_2 - e_1)}{e_1 e_2 (e_1+e_2)^2} r^2 \, (dr^2 + r^2 d\theta^2).
\ee
Changing coordinates to $u = r^2$, we have
\be
\frac{1}{\alpha'} \left.ds_\Sigma^2\right\rvert_{z_c} \approx \frac{(e_2 - e_1)}{e_1 e_2 (e_1+e_2)^2} \, (du^2 + 4 u^2 d\theta^2),
\ee
and the two-dimensional Ricci scalar $R_\Sigma$ on $\Sigma$ is
\be \label{eq: ricci BW2}
\sqrt{g_\Sigma}\, R_\Sigma = -4 \pi \delta(u),
\ee
i.e. there is a conical excess of $2 \pi$ at $z=z_c$ in $\Sigma$ along the $\AdS_2 \times S^2 \times S^4$ fiber. At the conical singularity, the remaining metric components take the values
\ie \label{eq: conical metrics BW2}
\frac{1}{\alpha'}\left.f_1^2\right\rvert_{z_c} &=\frac{(e_2+e_1)^2}{e_2-e_1} \\ 
\frac{1}{\alpha'}\left.f_2^2\right\rvert_{z_c} &=\frac{4e_1e_2}{e_2-e_1} \\
\frac{1}{\alpha'}\left.f_4^2\right\rvert_{z_c} &=e_2-e_1 .
\fe
It can also be found that the five-form flux $F_5$ is regular at $u=0$ in orthonormal coordinates and thus the combination $(F_5)^{2}_{MN} = (F_5)_{MPQRS} (F_5)_{N}^{PQRS}$ entering into the IIB supergravity equations must be regular at $z = z_c$.

The two-dimensional Ricci scalar $R_\Sigma$ on $\Sigma$ indicates that the two-cover geometry requires a codimension-2 source of negative tension at $z = z_c$ along the fiber $\AdS_2 \times S^2 \times S^4$, i.e. the dual operator on the $S^1$ conformal boundary must be a source of negative energy. Such negative Euclidean energy is typically required to construct Euclidean wormhole solutions. The on-shell action for the model of such a source and its connection to the matrix model free energy will be discussed in Section~\ref{sec: free energy}.

\subsubsection*{Two-to-one map to $\AdS_5 \times S^5$}

We can write down a map that makes manifest the fact that the Riemann surface $\Sigma$ and hence the 2-cover wormhole is a double cover of $\AdS_5 \times S^5$. Define the two-to-one transformation
\be \label{eq: two-to-one map}
w(z) = z - \frac{e_1 e_2}{z} \, , \qquad z_{\pm}(w) = \frac{w \pm \sqrt{w^2 + 4 e_1 e_2}}{2}
\ee
with ramification points at $w = \pm 2 i \sqrt{e_1 e_2}$, which correspond to images of the locations $z = \pm i \sqrt{e_1 e_2}$ of the conical singularities. Then the harmonic functions $h_{1,2}(w)$ become precisely those for the $\AdS_5 \times S^5$ geometry
\ie
h_1 &= \frac{\alpha'}{4}\sqrt{(e_2-e_1)^2 - w^2} \, + c.c. \, , \\
h_2 &= i \frac{\alpha'}{4} w + \, c.c. \,
\fe
where the branch cut is now at $w = [-\sqrt\lambda,\sqrt\lambda] = [-e_2 + e_1, e_2 - e_1]$ on the $w$-plane. The metric on the Riemann surface $\Sigma_w$ becomes the flat metric
\be
4 \rho^2(z,\bar{z}) d z d \bar{z} = \alpha' (e_2 - e_1) d w d \bar{w}  \, 
\ee
without the conical excess after the transformation.

The difference between the result of the map \eqref{eq: two-to-one map} and the $\adsS$ solution reviewed in Section \ref{vacuum AdS solution} is that, in the former, there are two branch cuts
\be
z = [-e_2, -e_1] \cup [e_1,e_2]
\ee
on the $z$-plane that each get mapped to a single cut
\be
\label{eqn: refer BW2 MM result}
w = [-\sqrt\lambda,\sqrt\lambda]
\ee
on the $w$-plane. In half-BPS Gaussian matrix models of $U(N)$ $\cN = 4$ SYM, a cut on the $w$-plane arises from the density of eigenvalues forming a Wigner semicircle. Later, we will identify the two overlapping cuts on the $w$-plane that result from the two-to-one map as the eigenvalue densities of two half-BPS Gaussian matrix models whose eigenvalues are bound together as a result of the insertion of a delta operator.

\subsection{Four-cover geometry}
\label{fourregions}

We now consider solutions with four poles on $\partial\Sigma$ at which the total metric asymptotes to $\AdS_5 \times S^5$. We refer to these geometries as a four-cover bubbling wormhole (BW$_4$).\footnote{We attempted but did not find an analytic expression for a three-cover geometry satisfying the regularity conditions.}

The regularity conditions eliminate simple solutions with fewer than four branch cuts. However, there exist a valid class of four-cut solutions, where each cut is placed in between the simple poles. If we fix the asymptotic $\AdS_5 \times S^5$ points to be located at $v=-a,0,a,\infty$, the solutions are given by
\begin{align}\label{fourboundarysolutions}
    & h_1 = \frac{\alpha'}{4} \sqrt{\frac{(e_2^2-z^2)(z^2-e_1^2)(z^2-a^4 e_1^{-2})(z^2-a^4 e_2^{-2})}{z^2(z-a)^2(z+a)^2}} \,+ \,\text{c.c.} \nonumber \\
    & h_2 = i \frac{\alpha'}{4} \left[z-\frac{a^2}{z}-\frac{(a^2-e_2^2)(a^2-e_1^2)}{2e_1e_2}\left(\frac{1}{z+a}+\frac{1}{z-a}\right)\right]\,+ \,\text{c.c.} ,
\end{align}
where $a>e_2>e_1>0$. The holomorphic part $\cA(z)$ of $h_1$ has branch cuts at
\be
z=[-a^2 e_1^{-1},-a^2 e_2^{-1}]\cup[-e_2,-e_1]\cup[e_1,e_2]\cup[a^2 e_2^{-1},a^2 e_1^{-1}].
\ee
There may exist a larger class of such solutions than the one written above, though symmetry makes the above particularly simple to write.

The metric functions, dilaton, and the fluxes are readily computed from the formulas in Section \ref{preliminarybubbling}. As before, the metric functions for $\AdS_2$, $S^2$, and $S^4$ do not vanish anywhere in the interior of $\Sigma$. $\AdS_2$ also does not vanish on $\partial\Sigma$, but $S^2$ or $S^4$ vanish in an alternating fashion on and off the branch cuts. The dilaton $\Phi = 2\phi$, and thus the string coupling $g_s$, takes the constant value
\be
g_s^2 = e^{4 \phi} = 1
\ee
throughout the bulk which we've normalized to be $1$.

There are three points in $\Sigma$ where $\rho^2(z,\bar{z})$ vanishes, located at
\ieg\label{four-cover conical points}
z=-ia \\
z = -\frac{i}{2}\left(\sqrt{\frac{(a^2-e_1^2)(a^2-e_2^2)}{e_1e_2}} \pm \sqrt{\frac{(a^2-e_1^2)(a^2-e_2^2)}{e_1 e_2} -4a^2}\right)
\feg
Each of these points correspond to conical excesses of $2 \pi$. When
\be
\frac{(a^2-e_1^2)(a^2-e_2^2)}{e_1e_2}=4a^2,
\ee
the total angle around the coalesced singularity is $8 \pi$ and the $\mathbb{Z}_2$-symmetries at the three conical points are promoted to a $\mathbb{Z}_4$-symmetry.

We can compute the charges corresponding to the flux across $S^5$ in each asymptotic $\AdS_5 \times S^5$ region just as in the two-boundary case. The D3 charges of all four regions are equal and
\be \label{eq: charge for four boundaries}
Q_{D3}^{(i)} = \frac{3 \pi}{2} \alpha'^2 \frac{ (e_2 - e_1)^2 (a^2 + e_1 e_2)^2}{e_1^2 e_2^2} \widehat{\text{Vol}}(S^4),
\ee
which yields
\ieg\label{4-cover WH lambda in terms of e2, e1, and a}
N^{(i)} = \frac{1}{4\pi} \frac{ (e_2 - e_1)^2 (a^2 + e_1 e_2)^2}{e_1^2 e_2^2} \\
\lambda = \frac{L_i^4}{\alpha'^2} = \frac{ (e_2 - e_1)^2 (a^2 + e_1 e_2)^2}{e_1^2 e_2^2}.
\feg

\subsubsection*{Four-to-one map to $\AdS_5 \times S^5$}

Define the four-to-one transformation
\be\label{four to one transform}
    w(z)=z-\frac{a^2}{z}-\frac{(a^2-e_2^2)(a^2-e_1^2)}{2e_1e_2}\left(\frac{1}{z+a}+\frac{1}{z-a}\right)
\ee
Under $w(z)$, the four-cover wormhole harmonic functions $h_1$ and $h_2$ becomes
\ie
h_1 &= \frac{\alpha'}{4}\sqrt{\frac{ (e_2 - e_1)^2 (a^2 + e_1 e_2)^2}{e_1^2 e_2^2} - w^2} \, +\,  \text{c.c.}, \\
h_2 &= i \frac{\alpha'}{4} w \,+\, \text{c.c.},
\fe
corresponding to the cut $w = [-\sqrt{\lambda}, \sqrt{\lambda}]$ on the $w$-plane. The locations \eqref{four-cover conical points} where $\rho^2(z,\bar{z})$ vanish correspond to the points at which the four-to-one map is degenerate. 

Under \eqref{four to one transform}, there are four branch cuts on the $z$-plane that each get mapped to the single cut $w = [-\sqrt{\lambda}, \sqrt{\lambda}]$ on the $w$-plane. In half-BPS Gaussian matrix models of $U(N)$ $\cN = 4$ SYM, we will identify the four overlapping cuts on the $w$-plane that result from the four-to-one map as the eigenvalue densities of four half-BPS Gaussian matrix models whose eigenvalues are bound together as a result of the insertion of delta operators.

\begin{figure}[t]
\centering
\includegraphics[width=150mm]{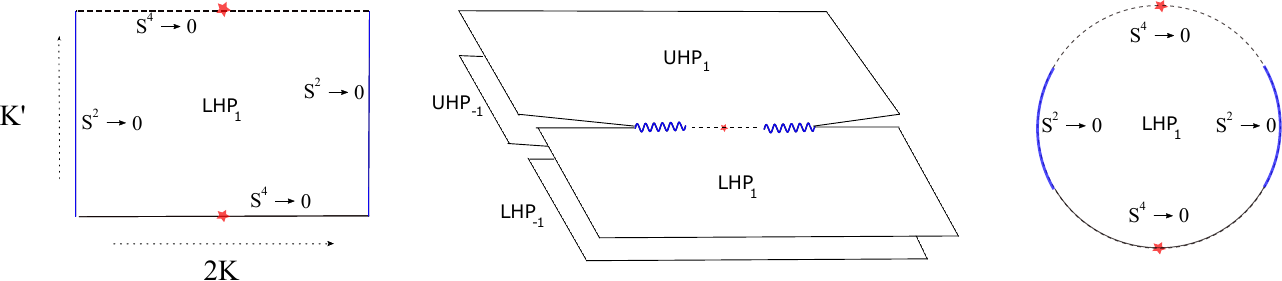}
\caption{The lower half-plane with two cuts and two singularities (marked points) is equivalent to the disk with two cuts and two marked points or to a rectangle with two marked points under the map $z = \sn(u,k)$. The harmonic function $h_2(u)$ vanishes at the edges of the rectangle that define the boundary of $\Sigma$. The blue cuts are the regions in which the $S^2$ vanishes and in the complementary boundary regions the $S^4$ vanishes.}
\label{fig:LHP}
\end{figure}

\subsection{Mapping $\Sigma$ to a regular polygon}

In this section, we perform an elliptic parametrisation of the functions involved that maps the $2n$-cut plane (or disk) with $2n$ singularities (including the one at infinity), to a regular polygon with $2n$ singularities.

Let us start with the simplest case of two cuts and two singularities. In this case, we found that the 2-cover wormhole is described by
\ie
h_1(z) &= \frac{\alpha'}{4} \sqrt{\frac{(e_{2}^2 - z^2)(z^2-e_{1}^2)}{z^2}} \,+\, {\rm c.c.} \\
h_2(z) &= i  \frac{\alpha'}{4} \left(z - \frac{e_{1} e_{2}}{z} \right) \,+\, {\rm c.c.} . 
\fe
The Riemann surface $\Sigma$ is the lower half-plane with two cuts and two singularities on its boundary. In this example there exists a bi-holomorphism between the doubly slit half-plane (or disk) with two singularities, to a rectangle with two singularities. One can explicitly construct this mapping using Jacobi's elliptic functions as follows.

We rescale $z = e_{1} \tilde{z} $ and define $\tilde{z} = \sn(u,k)$, with $k = e_{1}/e_{2}$. The harmonic functions become
\ie
h_1(u) &= \frac{\alpha'}{4} \frac{\cn(u,k) \dn(u,k)}{\sn(u,k)}  \,+\, {\rm c.c.} \\
h_2(u) &= i\frac{\alpha'}{4} \left(\sn(u,k) - \sn(u + i K',k) \right) \,+\, {\rm c.c.} . 
\fe
The lower half-plane $\Sigma$ is now mapped to the rectangle with four corners $u=-K, K, -K - i K', K - i K'$ that is a quadrant of the torus (the full plane with two cuts). The edges of the rectangle are the locii where $h_2(u) = 0$. In the $u$ plane there are no cuts and the elliptic functions have simple poles. In particular on the boundary
of $\Sigma$ we find again two singularities of $h_{1,2}$, one on $u=0$ and one on $u= -i K'$. These are the images of the singularities at $z = 0 , - \infty$. On the edges that contain the singularities, the $S^4$ part of the metric vanishes, while on the perpendicular edges without singularities, the $S^2$ part of the metric vanishes. The rectangle with two boundary marked points (singularities) has a $\mathbb{Z}_2 \times \mathbb{Z}_2 $ symmetry, the fixed point of this
symmetry is the point at the center $u = - i K'/2$. One can see a depiction of the geometry in Figure~\ref{fig:LHP}.

To check for singular points in the interior of $\Sigma$, that could lead to geometric bulk singularities, we compute
\ie
\partial_u h_1(u) &= \frac{\alpha'}{4} \frac{i (k \sn^2(u,k) -1)(k \sn^2(u,k) + 1)}{\sn^2(u,k)} \, , \\
\partial_u h_2(u) &= i \frac{\alpha'}{4} \cn(u,k) \dn(u,k) \frac{1 + k \sn^2(u,k) }{k \sn^2(u,k)}  \, . 
\fe
We observe that in these coordinates that form a double cover of the plane, the point at $z = - i \sqrt{e_{1} e_{2}}$ is mapped to $u = - i K'/2$, that is the fixed point of the discrete symmetries. At this point we find that both $\partial_u h_1 = \partial_u h_2 = 0$ vanish together and we expect a possible conical excess singularity according to the analysis of section~\ref{tworegions}. The other possible zeroes of $\partial_u h_2(u)$ are exactly at the corners of the rectangle (that are the images of the endpoints of the cuts) and do not lead to singularities in the metric functions. 

The Riemann surface $\Sigma$, described by the square in Figure~\ref{fig:LHP}, has the following metric in this coordinate system
\be\label{sigmametric}
d \Sigma^2 = 4 \rho^2(u) d u d \bar{u} \propto \frac{  |1 + k \sn^2(u,k)|^2}{|\sn(u,k)|^2 } |d u|^2 
\ee
In order to check for the presence of a conical excess in the center of the rectangle, we expand $u = - i K'/2 + r e^{i \theta}$ we find 
\be
d \Sigma^2 \propto r^2 ( dr^2 + r^2 d \theta^2 ) .
\ee
We observe again the presence of a conical excess of $2 \pi$ in Section \ref{tworegions}.

This mapping can be generalised for geometries with more than two boundaries using the appropriate hyperelliptic functions. In particular the disk has $2n$ singularities that are separated by $2n$ cuts and this more general geometry can be mapped to a regular canonical polygon with $4n$ edges and $2n$ singularities (marked points). The regular polygon manifests both the cyclic (rotational) symmetry and the reflection $\mathbb{Z}_2$-symmetries of the geometry.

\section{Review: Wilson loops and localization in $\mathcal{N}=4$ SYM}\label{WilsonLoopreview}

The expectation value of a half-BPS Wilson loop in a representation $R$ of the $U(N)$ $\cN = 4$ super Yang-Mills on $S^4$ can be represented by a Hermitian matrix integral \cite{Semenoff:2001xp,Pestun:2007rz}
\be\label{singlehalfBPS}
\langle W_R \rangle = \frac{1}{Z} \int \mathcal{D} M \,  e^{- \frac{2 N}{\lambda} \tr M^2} \tr_R \left( e^{M} \right) \, .
\ee
It is also possible to compute correlators of several half-BPS Wilson loops by inserting multiple traces, each transforming in a representation $R_i$, into the matrix integral \eqref{singlehalfBPS}.

Holographic duals of such Wilson loops have been a subject of much investigation. A Wilson loop in a small representation with $O(1)$ boxes is dual to a fundamental string on $\AdS_2$ \cite{Maldacena:1998im,Rey:1998ik,Drukker:1999zq,Zarembo:2002an}. An (anti‑)symmetric representation with $O(N)$ boxes is dual to a D3-brane on $\AdS_2 \times S^2$ or a D5-brane on $\AdS_2 \times S^4$ \cite{Drukker:2005kx,Hartnoll:2006is,Yamaguchi:2006tq,Gomis:2006im,Gomis:2006sb}. A Wilson loop in a large representation with $O(N^2)$ boxes corresponds to a backreacted half-BPS gravity solution described by a metric of the form \eqref{eq: metric ansatz} that contains the $\AdS_2 \times S^2 \times S^4$ factor \cite{Yamaguchi:2006te,DHoker:2007mci,Okuda:2008px}. Remarkably, the planar resolvent $\omega(z)$, which can be split into a ``classical'' piece and a ``quantum'' spectral curve piece, completely determines the form of the dual geometry, as we shall soon describe.

\begin{figure}[t]
\centering
\includegraphics[width=100mm]{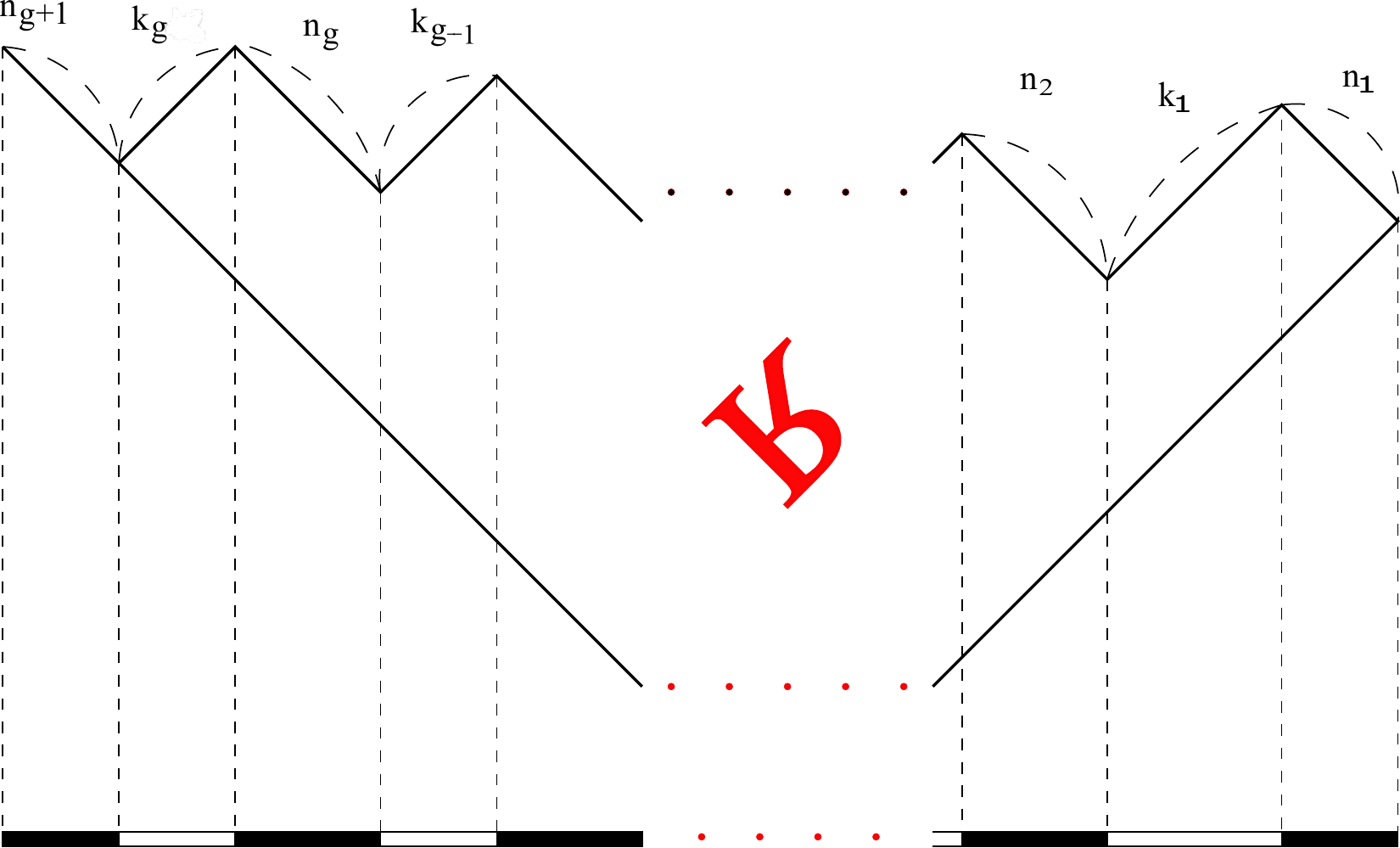}
\caption{The rotated and flipped Young diagram corresponding to the representation $R$. It is described by a collection of rectangular blocks of size $\{n_I, K_I\}_{I=1}^{g+1}$, specifying the number of rows and columns. $n_{g+1}$ is fixed by $n_{g+1} + \sum_{I=1}^g n_I = N$, and $K_{g+1}=0$. Once projected onto the real line, one produces a Maya diagram consisting of black and white lines, whose sizes depend on $n_I$ and $k_I = K_I - K_{I+1}$ respectively. These lines correspond to the cuts of the matrix model spectral curve $y(z)$, which in turn dictates the form and properties of the dual supergravity geometry. The parameter $g$ fixes the number of cuts (the genus of the spectral curve $y(z)$) and the shape of the associated Riemann surface.}
\label{fig:Young}
\end{figure}

Let us now explain how to derive the resolvent and spectral curve for a single half-BPS Wilson loop in an irreducible representation $R$. Starting from the matrix integral \eqref{singlehalfBPS}, the expectation value is given by the following integral over the eigenvalues $m_i$ of the Hermitian matrix $M$
\be\label{Wilson loop vev from MM}
\langle W_R \rangle = \frac{1}{Z} \int \prod_{i=1}^N d m_i \, \Delta^2(m)\, e^{- \frac{2 N}{\lambda} \sum_i m_i^2} \tr_R \left( e^{m} \right),
\ee
where $\Delta(m)$ is the Vandermonde determinant. The character associated to an irrep $R$ is given by
\be
\tr_R \left( e^m \right) \, = \, \chi_R \left( e^m \right) \, = \, \frac{\det_{i j} e^{m_i (\n_j + N - j)}}{\Delta (e^{m})} \, ,
\ee
where $\n_i$ counts the number of boxes on each row of the Young diagram associated to $R$.

By using the antisymmetric properties of determinants, one can simplify the expression for the expectation value of the Wilson loop into a product of diagonal terms
\be\label{oneloopeigen}
\langle W_R \rangle = \frac{N!}{Z} \int \prod_{i=1}^N d m_i \, \Delta^2(m) \, e^{- \frac{2 N}{\lambda} \sum_i m_i^2} \, \frac{ \prod_{i=1}^N e^{m_i h_i}}{\Delta (e^{m})} \, , \quad h_i = \n_i + N - i \, ,
\ee
where we also defined the shifted highest weights $h_i \geq 0$ (strictly decreasing). 
We would like to understand the form of the dual geometry in the large $N$ limit, when the representations are large and backreact on the background. In this limit, the Young diagram has $O(N^2)$ boxes and therefore the highest weights are $\n_i \sim O(N)$. A useful approach to analyzing a ``heavy'' Wilson loop in a fixed representation $R$ involves decomposing the representation into $g$ blocks, each containing $n_I$ rows of length $K_I$. This splits the range $i \in [1,N]$ into $g+1$ segments labeled by $I \in [1,g+1]$, each of length $n_I$, such that $n_{g+1}= N - \sum_{I=1}^g n_I$. $K_I = \sum_{J=I}^g k_J$ for $I\in[1,g]$ and $K_{g+1}=0$, where $k_J = K_J - K_{J+1}$ labels the difference in the number of columns between consecutive blocks. See Figure~\ref{fig:Young} for details.

Once the matrix model has been reduced to integrals over its eigenvalues as in \eqref{oneloopeigen}, the most basic manipulation is to solve the saddle point equations
\be
\int \prod_{i=1}^N d m_i  \, e^{- S_\text{eff}(m_i) } \, ,  \qquad \frac{\delta S_\text{eff}(m_i)}{\delta m_i} = 0 \, .
\ee
The effective action $S_\text{eff}(m_i)$ contains both the potential $V_\text{cl.}(m_i)$ that was originally in the exponent (the ``classical'' part), as well as any terms from the measure after exponentiation (the ``quantum'' part). For such an eigenvalue integral, the most basic quantities to compute are the so-called resolvent $\omega(z)$ and the density of eigenvalues $\rho(z)$. The support $\supp \rho$ is a branch cut $\mathcal{C}$ of $\omega(z)$ on the real axis.
\be
\omega(z) = \int_{\mathcal{C}} d z' \, \frac{\rho(z')}{z-z'} \, , \qquad \rho(z) = \frac{1}{N} \sum_{i=1}^N \delta(z - m_i) \, , \qquad \int_{\mathcal{C}} d z \, \rho(z) = 1 \, .
\ee
The resolvent $\omega(z)$, which encapsulates the solution of the matrix model saddle point equations, can be most easily determined in the large $N$ limit (``planar'' resolvent). Similarly to the effective action, it can be split into a ``classical'' and ``quantum'' piece---called the spectral curve $y(z)$---from which one can determine the density of eigenvalues:
\be\label{splittingres}
2 \omega(z) = V_\text{cl.}'(z) - y(z)  \, , \qquad \rho(z) =  \frac{1}{2 \pi} \Im y(z) \, , \quad z \in \mathcal{C} \, .
\ee

In \cite{Okuda:2008px}, it was found that there is a direct relation between the matrix model resolvent of the integral \eqref{oneloopeigen} and the harmonic functions $h_1 = \mathcal{A} + \bar{\mathcal{A}}$ and $ h_2 = \mathcal{B} + \bar{\mathcal{B}}$ that specify the dual bubbling geometry, see Section \ref{preliminarybubbling}. This connection comes after splitting the matrix model resolvent in the classical and quantum (spectral curve piece) as follows 
\be\label{matchingrelation}
  \mathcal{B} = i \frac{\pi \alpha' g_s N}{4} V_\text{cl.}'(z)  \, , \quad \mathcal{A} = i \frac{\pi \alpha' N}{4} y(z) \, .
\ee
The ``classical piece'' of the matrix model resolvent stems from the matrix model potential. For the Gaussian matrix model, it is simply $V_\text{cl.}(z) = 2 z^2/\lambda$, and is the one dictating the boundary $\partial \Sigma$ of the Riemann surface $\Sigma$. On the other hand, the spectral curve determines the cuts on $\partial \Sigma$ and the physical properties (such as the metric and the fluxes) of the dual geometry. In our work we propose that~\eqref{matchingrelation} is the correct generalization of the results for the Gaussian matrix model, even in the multi-boundary case. We shall explicitly verify that this is the case in section~\ref{MMbubbling}.

Let us momentarily go back to the Gaussian matrix model and describe how this works in the simplest example when we replace the Wilson loop operator \eqref{oneloopeigen} with the identity operator, i.e. trivial representation. We have a Gaussian matrix model with an effective action
\be\label{single-matrix Gaussian model}
S_\text{eff}(m_i) = - \frac{2 N}{\lambda} \sum_i m_i^2 + \sum_{i \neq j } \log |m_i - m_j| \, ,
\ee
and saddle point equations
\be\label{gaussiansaddleeqn}
\frac{4 N}{\lambda}  m_i = \sum_{i \neq j } \frac{2}{m_i - m_j} \, ,
\ee
which at large $N$ can be written using the density of eigenvalues and the resolvent as
\be
\frac{4}{\lambda}  z = 2 \int d z' \, \frac{\rho(z')}{z - z'} = \omega(z + i\epsilon) + \omega(z - i \epsilon) \, .
\ee
The resolvent can then be determined to be
\be\label{1-cut Wigner resolvent}
\omega(z) = \frac{2}{\lambda} z - \frac{2}{\lambda} \sqrt{z^2 - \lambda } \, ,
\ee
from which one can find the spectral curve and density of eigenvalues 
\bea\label{Wigner semicircle}
    y(z) = \frac{4}{\lambda} \sqrt{z^2 - \lambda}, \quad \rho(z) = \frac{2}{\lambda \pi} \sqrt{\lambda - z^2}
\eea 
using \eqref{splittingres}, and then the harmonic functions via \eqref{matchingrelation} and \eqref{holomorphicmetricfunctions}. This density is known as the Wigner semicircle. The dual geometry stemming from this resolvent can be seen to correspond to $\AdS_5 \times S^5$, see Section \ref{vacuum AdS solution}.

This analysis can be extended for general Wilson loops in fixed irreps $R$. In these cases, the role of the shifted highest weights is that of a linear source coupled to the eigenvalues via the term $m_i h_i$ in the effective action, driving the eigenvalues away from the origin. Instead of \eqref{gaussiansaddleeqn}, the relevant saddle point equation now becomes for the eigenvalues of the $\mathcal{I}_I$ block
\be\label{gaussiansaddleeqn2}
\frac{4 N}{\lambda} \left(  m_i - \frac{K_I \lambda}{4 N} \right) = \sum_{i \neq j } \frac{2}{m_i - m_j} \, , \qquad m_i \in \mathcal{I}_I \, , \quad I = 1, ... g+1 \, ,
\ee
see above discussion and Figure~\ref{fig:Young} for the definition of $K_I$. This leads to a $(g+1)$-cut solution, with several cuts being displaced from the origin due to the sources $K_I$ \cite{Okuda:2008px}. The resolvent in general is expressed in terms of (hyper) elliptic functions. Assuming though that the distances between the cuts are large, so that the eigenvalues between different intervals do not interact among themselves (valid at large $\lambda$ when the Gaussian potential becomes very wide), one can approximate the multi-cut solution by a superposition of Wigner-semicircle densities (here $n_I, K_I \sim O(N)$, so $b_I, c_I$ are $O(1)$):
\be\label{Wignersuperpos}
\rho(z) \approx \frac{2}{\pi \lambda} \sum_{I=1}^{g+1} \sqrt{\lambda b_I - \left(z - c_I \right)^2 } \, , \qquad b_I = \frac{n_I}{N} \, , \quad c_I = \frac{K_I \lambda}{4 N} \, . 
\ee
This approximation is consistent at large $\lambda$, since the size of the cuts is $O(\sqrt{\lambda})$, while the distance between their centers is $O(\lambda)$. This concludes our review of the multi-cut solutions for Wilson loops in large but fixed irreducible representations $R$.

In contrast to the case of irreps $R$, holographic realizations of BPS Wilson loops in large reducible representations or sums over representations, such as the one appearing in \eqref{correlatedloops}, have not been widely investigated. While a generic sum over all representations would be difficult to study, we focus on a class of operators where it is possible to perform the sum over representations exactly as in the delta operator \eqref{heavydeterminant}. It is tractable in these examples to analyze and solve the saddle point equations of the resulting multi-matrix model.

\section{Matrix model for bubbling wormholes}
\label{MMbubbling}

\subsection{Two-cover spectral curve from a delta operator}\label{twomatrixequations}

We now construct a two-matrix model whose spectral curve describes the two-cover BW$_2$ in Section~\ref{tworegions}. We begin with two decoupled Gaussian matrix models
\bea\label{decoupled 2-Gaussian matrix model}
     Z_1 Z_2 = \int\mathcal{D}M_1\mathcal{D}M_2\,e^{-\frac{2N}{\lambda}\Tr M_1^2-\frac{2N}{\lambda}\Tr M_2^2}
\eea 
which correspond to two decoupled $U(N)$ $\mathcal{N}=4$ super Yang-Mills theories SYM$_{1,2}$.

Let $\mathbf{K}$ denote the rectangular Young diagram $(K^N)$ with $N$ rows and $K\sim O(N)$ columns.\footnote{It is useful to note that $\Tr_{\mathbf{K}}(e^M)=\det^K (e^M)$, which is the product of the eigenvalues of $e^M$, each raised to the $K$-th power.} The observables $\Tr_{\bf K}(e^{M_1})$ and $\Tr_{\bf K}(e^{-M_2})$ in the two half-BPS matrix models, which correspond to BPS Wilson loops of opposite orientations in irrep ${\bf K}$, are dual to two disconnected bubbling geometries of a simple type. In the eigenvalue description where we denote the eigenvalues of $M_1$ and $M_2$ as $\{x_i\}_{i=1}^N$ and $\{y_i\}_{i=1}^N$, respectively, the effect of the insertions $\Tr_{\bf K}(e^{M_1})$ and $\Tr_{\bf K}(e^{-M_2})$ is to shift the center of the Gaussian potential by an amount proportional to $\lambda$ in each direction:\footnote{Our construction can also be interpreted in terms of two decoupled SYM$_{1,2}$ whose scalars $\Phi_0^{(1)}$ and $\Phi_0^{(2)}$ of \eqref{def. of half-BPS Wilson loop in N=4 SYM} are given vevs of order $\lambda$ in the 't Hooft limit where $N \to \infty$ as $\lambda$ is large but fixed.}
\ie\label{eq: eigenvalue K}
&\la \Tr_{\bf K}(e^{M_1}) \ra_1 \la \Tr_{\bf K}(e^{-M_2}) \ra_2 \\
&= \frac{e^{K^2\lambda/8}}{Z_1} \int \prod_i d x_i \, \Delta^2(x) \, e^{ - \frac{2N}{\lambda} \sum_{i} \left(x_i - \frac{K \lambda}{4N}\right)^2 } \cdot \frac{e^{K^2\lambda/8}}{Z_2} \int \prod_i d y_i \, \Delta^2(y) \, e^{ - \frac{2N}{\lambda} \sum_{i} \left(y_i + \frac{K \lambda}{4N}\right)^2 }.
\fe
We will work on the background given by these operators dual to ``simple'' disconnected bubbling solutions. Let us denote the expectation value of an operator $\cO(M_1, M_2)$ taken in the background of $\Tr_{\bf K}(e^{M_1})$ and $\Tr_{\bf K}(e^{-M_2})$ as
\ie \label{eq: 12K expectation}
\la \cO(M_1, M_2) \ra_{12}^{\bf K} = \frac{e^{K^2\lambda/4}}{Z_1 Z_2} \int \prod_i d x_i  d y_i \, \Delta^2(x) \Delta^2(y) \, e^{ - \frac{2N}{\lambda} \sum_{i} \left(x_i - \frac{K \lambda}{4N}\right)^2 - \frac{2N}{\lambda} \sum_{i} \left(y_i + \frac{K \lambda}{4N}\right)^2 } \cO(x,y).  
\fe
We find that the insertion $\la \, \widehat\delta_{12} \, \ra_{12}^{\bf K}$ of the supersymmetric delta operator
\ie \label{eq: delta operator}
\widehat\delta_{12} &= \frac{\det ( \mathds{1} \otimes \mathds{1} + e^{M_1} \otimes e^{-M_2} )}{\det ( \mathds{1} \otimes \mathds{1} - e^{M_1} \otimes e^{-M_2} )} \\
&= \left( \sum_R \Tr_R (e^{M_1}) \, \Tr_R (e^{-M_2})\right) \left( \sum_R \Tr_{R^T} (e^{M_1}) \, \Tr_R (e^{-M_2}) \right)
\fe
on this background has the effect of identifying the eigenvalue densities on the two disconnected spectral curves of $\la \Tr_{\bf K}(e^{M_1}) \ra_1$ and $\la \Tr_{\bf K}(e^{-M_2}) \ra_2$. The result of the identification will coincide with the spectral curve of the two-cover bubbling wormhole BW$_2$.

Let us study the effect of the delta operator $\widehat\delta_{12}$ on the matrix models. In terms of eigenvalues, its expectation value is
\be\label{2-cover WH matrix model eigenvalue basis}
\la \, \widehat\delta_{12} \, \ra_{12}^{\bf K} \propto \int \prod_i d x_i d y_i \, \Delta^2(x) \Delta^2(y) \, e^{ - \frac{2N}{\lambda} \sum_{i} \left(x_i - \frac{K \lambda}{4N}\right)^2 - \frac{2N}{\lambda} \sum_{i} \left(y_i + \frac{K \lambda}{4N}\right)^2 } \prod_{i,j} \frac{1+ e^{x_i-y_j}}{1- e^{x_i-y_j}}
\ee 
The finite-$N$ saddle point equations of this two-matrix model are
\bea\label{saddlepequationsmixed}
\frac{4N}{\lambda}\left(x_i-\frac{K\lambda}{4N}\right)=\sum_{j(\neq i)}^N\frac{2}{x_i-x_j} - \sum_{j=1}^N \frac{1}{\sinh(x_i-y_j)} \, ,\nn \\
\frac{4N}{\lambda}\left(y_i+\frac{K\lambda}{4N}\right)=\sum_{j(\neq i)}^N\frac{2}{y_i-y_j} - \sum_{j=1}^N \frac{1}{\sinh(y_i-x_j)} \, .
\eea
These equations describe a two-flavor gas of eigenvalues with $x$-$x$ and $y$-$y$ repulsion and an $x$-$y$ attraction. Besides the linear force from the Gaussian potential, there is a large external force of order $N$ pulling the $x$-eigenvalues to the right, while an external force of the same magnitude pulls the $y$-eigenvalues to the left. In the absence of an $x$-$y$ cross-interaction, these external forces induce opposite vevs $\pm \frac{K\lambda}{4N}\sim O(\lambda)$ for $M_1$ and $M_2$.

An important property of eqns.~\eqref{saddlepequationsmixed} is that they admit a certain $x$-$y$ bound-state solution at large 't Hooft coupling $\lambda$. Let us first perform a scaling analysis for the single-matrix Gaussian model \eqref{single-matrix Gaussian model} and then extend it to our two-matrix model \eqref{2-cover WH matrix model eigenvalue basis} to understand the bound-state configuration.

Suppose that the eigenvalues of the Gaussian matrix model \eqref{single-matrix Gaussian model} are distributed along a segment of length $L$. Given $N$-eigenvalues, the typical microscopic spacing between nearest-neighbor eigenvalues is of order $\Delta \sim L/N$\footnote{Near the edges of the spectrum the spacing changes but we neglect such effects in what follows.}. The typical distance between a pair of eigenvalues $x_i$ and $x_j$ is then $x_i - x_j \sim (i-j) \Delta$. Therefore, the force-balancing condition \eqref{gaussiansaddleeqn} for $x_i \sim L$ implies the following scaling relation between $L$ and $\lambda$:
\be
 \frac{N}{\lambda} L \sim \frac{4 N}{\lambda} x_i = \sum_{j (\neq i)}^N \frac{2}{x_i - x_j} \sim \frac{1}{\Delta} \sum_{j (\neq i)}^N \frac{1}{i-j} \sim \frac{N}{L}  \quad \Rightarrow \quad \Delta \sim \frac{L}{N} \sim \frac{\sqrt{\lambda}}{N}
\ee
in agreement with the Wigner semicircle solution \eqref{Wigner semicircle}. 

We consider the following pairwise bound-state ansatz for the saddle equations \eqref{saddlepequationsmixed}
\bea\label{2-cover bound-state}
    (y_1\lesssim x_1) < (y_2 \lesssim x_2) < \cdots < (y_N \lesssim x_N)
\eea 
More precisely, we require that for all diagonal $i$, we have
\bea\label{2-cover bound-state condition 1}
    x_i-y_i \approx \frac{1}{K} \sim \frac{1}{N}
\eea 
and for all $(i,j)$ pairs where $i \neq j$, we have
\bea 
    x_i - x_j \approx y_i - y_j \sim (i-j) \frac{\sqrt{\lambda}}{N}.
\eea 
In the semiclassical regime $1 \ll \lambda \ll N$, we have that, for all $i \neq j$,
\bea\label{2-cover bound-state condition}
    x_i-y_i  \ll x_i - x_j \approx y_i - y_j .
\eea 
That is, the size $\delta \equiv x_i - y_j$ of any bound state is parametrically smaller than the average spacing $\Delta \equiv x_{i+1} - x_i \approx y_{i+1} - y_i$ between the neighboring bound states. The ratio $\delta / \Delta \sim 1/\sqrt{\lambda}$ goes to zero in the limit of large 't Hooft coupling $\lambda$.

An immediate consequence of \eqref{2-cover bound-state condition 1} is that the constant external forces balance with the diagonal cross-terms in \eqref{saddlepequationsmixed} (note that this balancing property does not depend on $\lambda$):
\be\label{bound-state config 2-cover}
    \frac{1}{\sinh(x_i-y_i)} \approx \frac{1}{x_i-y_i} \approx K .
\ee
Therefore, the saddle point equations \eqref{saddlepequationsmixed} reduce to 
\ie\label{saddlepequations still coupled}
\frac{4N}{\lambda}x_i &\approx \sum_{j(\neq i)}^N\frac{2}{x_i-x_j} - \sum_{j(\neq i)}^N \frac{1}{\sinh(x_i-y_j)} \, , \\
\frac{4N}{\lambda}y_i &\approx \sum_{j(\neq i)}^N\frac{2}{y_i-y_j} - \sum_{j(\neq i)}^N \frac{1}{\sinh(y_i-x_j)} \, ,
\fe
with the proviso that $x_i \approx y_i$ has formed a bound pair for all $i$. The equations~\eqref{saddlepequations still coupled} simplify significantly in the limit of large $\lambda$. Let us argue that in this limit the off-diagonal cross-terms can be dropped
\bea
    \sum_{j(\neq i)}^N \frac{1}{\sinh(x_i-y_j)} \ll \sum_{j(\neq i)}^N\frac{2}{x_i-x_j} \approx \sum_{j(\neq i)}^N\frac{2}{y_i-y_j}.
\eea 
When $x_i - y_j \ll 1$, we can approximate $1/\sinh (x_i - y_j) \approx 1/(x_i - y_j)$, while for $x_i - y_j \gg 1$, one can approximate $1/\sinh (x_i - y_j) \approx 0 $. The transition between these two regimes happens when the inter-eigenvalue distance $x_i - y_j$ is
\bea
    x_i - y_j \sim (i - j) \Delta \sim O(1).
\eea 
Since the width of the bound-state eigenvalue distribution is of order $\sqrt{\lambda}$, any $O(1)$ segment of this distribution away from the edges is approximately uniform in the large $\lambda$ limit. Consequently, for a given $x_i$, the $ 1/\sinh (x_i - y_j)$ terms cancel pairwise within an $O(1)$ neighbourhood of $x_i$, because the distribution of $y_j$ is approximately symmetric to the left and right of $x_i$ in this neighbourhood. In this approximation, the off-diagonal cross-terms can be neglected when solving the saddle point equations for the bound-state configuration.

The saddle point equations \eqref{saddlepequations still coupled} on the bound-state configuration then reduce to the simple equations
\be\label{saddlepequationsdecoupled}
\frac{4N}{\lambda}x_i\approx \sum_{j(\neq i)}^N\frac{2}{x_i-x_j} , \quad \frac{4N}{\lambda}y_i\approx \sum_{j(\neq i)}^N\frac{2}{y_i-y_j}.
\ee
whose solutions are Wigner semicircles
\be\label{solutions to saddlepequationsdecoupled}
    \rho(x)=\frac{2}{\lambda \pi}\sqrt{\lambda -x^2},\quad \rho(y)=\frac{2}{\lambda \pi}\sqrt{\lambda -y^2}.
\ee
This provides an a posterori justification for our ansatz of an order $\sqrt{\lambda}/N$ spacing between the neighboring $x$-eigenvalues (or $y$-eigenvalues). The only remnant of the presence of the bound pairs $x_i$-$y_i$ eigenvalues is that the two distributions are overlapping and identified, recovering the result of the 2-1 map of the two-cover bubbling wormhole in Section \ref{tworegions}. In particular, see the discussion around \eqref{eqn: refer BW2 MM result}.

An alternative way to see that the bound-state solution \eqref{solutions to saddlepequationsdecoupled} reproduces the branch cuts $[-e_2,-e_1]\cup[e_1,e_2]$ and the simple poles at $0$ and $\infty$ of the two-cover bubbling wormhole $z$-plane (see Figure~\ref{fig:DoubleCover}) is via the coordinate transformation
\ie\label{1-2 coordinate transform} 
    z(x) = \frac{x - \sqrt{x^2 + 4 e_1 e_2}}{2}, \quad z(y) = \frac{y + \sqrt{y^2 + 4 e_1 e_2}}{2}
\fe 
Using \eqref{2-cover WH lambda = e2 - e1}, we can show that the branch cut $[-\sqrt{\lambda},\sqrt{\lambda}]$ and the simple pole at $\infty$ in the $x$-plane map to the branch cut $[-e_2,-e_1]$ and the simple pole at $0$ in the $z$-plane, while the branch cut $[-\sqrt{\lambda},\sqrt{\lambda}]$ and the simple pole at $\infty$ in the $y$-plane map to the branch cut $[e_1,e_2]$ and the simple pole at $\infty$ in the $z$-plane. Moreover, the bound-state eigenvalue density \eqref{solutions to saddlepequationsdecoupled} in the $z$-variable \eqref{1-2 coordinate transform} is
\ie 
    \rho(z) = \frac{2}{\lambda \pi} \sqrt{\frac{(e_2^2 - z^2)(z^2 - e_1^2)}{z^2}}
\fe 
which can be found by taking the discontinuity (see \eqref{splittingres}) of the following resolvent
\ie\label{2-cover resolvent in z-variable} 
    \omega (z) = \frac{2 }{\lambda} \left(z-\frac{e_1 e_2}{z}-\frac{1}{z}\sqrt{(z^2-e_1^2)(z^2-e_2^2)}\right)
\fe
across its branch cuts. It is interesting to note that \eqref{2-cover resolvent in z-variable} is the resolvent of the two-cover Gaussian–Penner matrix model \eqref{eqn: 2-cover GP MM}, which is discussed in more detail in Appendix~\ref{appendix: Gaussian-Penner model}. Applying the relation \eqref{matchingrelation} to the two-cover Gaussian–Penner resolvent \eqref{2-cover resolvent in z-variable} yields precisely the harmonic functions $h_{1,2}$ of BW$_2$. We can therefore regard this two-cover Gaussian–Penner matrix model \eqref{eqn: 2-cover GP MM} as an effective matrix model description of BW$_2$. By contrast, the two-matrix model \eqref{2-cover WH matrix model eigenvalue basis}, together with the coordinate transform \eqref{1-2 coordinate transform}, constitutes our proposal for how BW$_2$ microscopically emerges from two originally decoupled Gaussian matrix models via the delta operator \eqref{eq: delta operator}, which we interpret as an entangled sum of Wilson loops.

\begin{figure}
    \centering
    \includegraphics[width=0.7\linewidth]{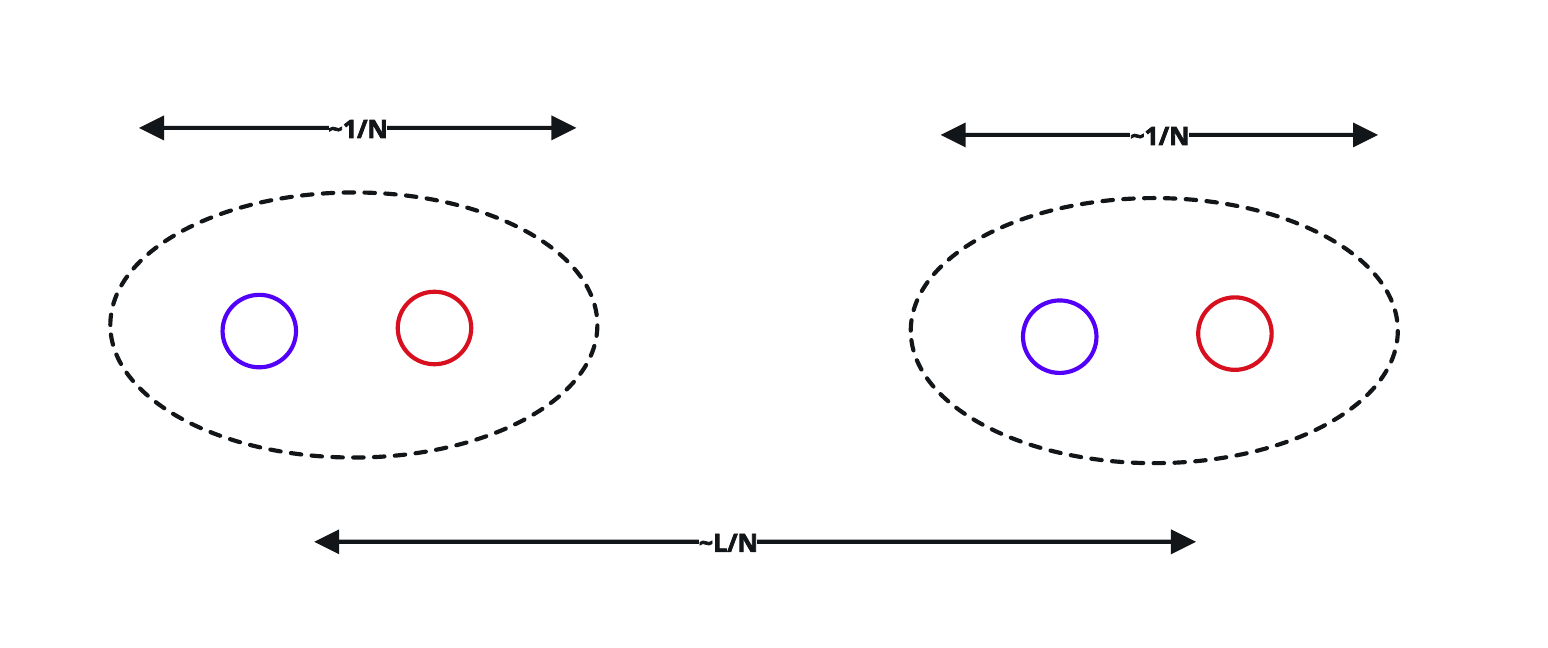}
    \caption{The formation of bound pairs between the two species of eigenvalues. The size of the bound-pair scales as $ 1/N$, when the distance between different pairs scales as $L/N \sim \sqrt{\lambda}/N$ (The $N$ bound-pairs get distributed on a segment of size $L$).}
    \label{fig:BoundPairs}
\end{figure}

Lastly, we comment why it was necessary to consider a supersymmetric version of the delta operator coupling the two matrix models. A ``bosonic'' delta operator
\be
\frac{1}{\det ( \mathds{1} \otimes \mathds{1} - e^{M_1} \otimes e^{-M_2} )} = \prod_{i,j}\frac{1}{1-e^{x_i-y_j}} \, ,
\ee
on its own would have contributed to the saddle equations \eqref{saddlepequationsmixed} the term
\be
-\sum_{j=1}^N\frac{e^{x_i-y_j}}{e^{x_i-y_j}-1}
\ee
instead of the ``supersymmetric'' one $-\sum_{j=1}^N\frac{1}{\sinh(x_i-y_j)}$. In contrast to the latter, the bosonic operator decays to 0 as $x-y\to-\infty$ but approaches the constant value $-1$ as $x-y\to+\infty$. This would be problematic for forming a bound-state configuration.

\subsection{Four-cover spectral curve from cyclic deltas}\label{fourcovercyclicdeltas}

We now generalize our matrix model considerations for BW$_2$ to BW$_4$ analyzed in Section~\ref{fourregions}. The idea is to consider four half-BPS matrix models that interact cyclically via the delta operators \eqref{eq: delta operator}.

We begin with four decoupled half-BPS matrix models in the background of the insertions
\be \label{eq: four background K}
\Tr_{\bf K_1} (e^{M_1}) \, \Tr_{\bf K_1} (e^{-M_2}) \, \Tr_{\bf K_2} (e^{M_3}) \, \Tr_{\bf K_2} (e^{-M_4})
\ee
each dual to bubbling geometries labelled by Young diagrams $\bf K_{1}$ or $\bf K_{2}$. Here, $\bf K_{1,2}$ denotes a rectangular Young diagram $(K_{1,2})^N$ with $N$ rows and $K_{1,2} \sim O(N)$ columns. We will find that the existence of a bound-state solution requires $K_1 - K_2$ to scale parametrically with $K_1 + K_2$, and we take $K_1 > K_2$ without loss of generality.

We again work on the background given by operators \eqref{eq: four background K} dual to disconnected bubbling solutions. Let $\{x_i\}_{i=1}^N$, $\{y_i\}_{i=1}^N$, $\{u_i\}_{i=1}^N$, and $\{v_i\}_{i=1}^N$ denote the eigenvalues of $M_1$, $M_2$, $M_3$, and $M_4$ respectively. Let us denote the expectation value of an operator $\cO(M_1,M_2,M_3,M_4)$ taken in the background of \eqref{eq: four background K} as
\ie \label{eq: 1234K expectation}
&\la \cO (M_1,M_2,M_3,M_4) \ra_{1234}^{\bf K_{1,2}} = \frac{e^{(K_1^2 + K_2^2)\lambda/4}}{Z_1 Z_2 Z_3 Z_4} \\
&\times \int \prod_i d x_i d y_i d u_i  d v_i \ \Delta^2(x)  \Delta^2(y)  \Delta^2(u) \Delta^2(v) \, e^{ - \frac{2N}{\lambda} \sum_{i} \left(x_i - \frac{K_1 \lambda}{4N}\right)^2 - \frac{2N}{\lambda} \sum_{i} \left(y_i + \frac{K_1 \lambda}{4N}\right)^2} \\
&\times \ e^{ - \frac{2N}{\lambda} \sum_{i} \left(u_i - \frac{K_2 \lambda}{4N}\right)^2 - \frac{2N}{\lambda} \sum_{i} \left(v_i + \frac{K_2 \lambda}{4N}\right)^2 } \cO(x,y,u,v)  
\fe
We find that the cyclic insertion of delta operators \eqref{eq: delta operator}
\be \label{eq: 4 cyclic deltas}
\la \, \widehat\delta_{13} \, \widehat\delta_{32} \, \widehat\delta_{24} \, \widehat\delta_{41} \, \ra_{1234}^{\bf K_{1,2}}
\ee
of the type $M_1$-$M_3$-$M_2$-$M_4$-$M_1$-$\cdots$ on this background has the effect of identifying the eigenvalue densities on the four disconnected spectral curves of \eqref{eq: four background K}. The result of the identification will coincide with the spectral curve of BW$_4$.

Let us study the effect of the cyclic insertion \eqref{eq: 4 cyclic deltas} on four half-BPS matrix models
\ie \label{4-cover MM eigenvalue basis}
&\la \, \widehat\delta_{13} \, \widehat\delta_{32} \, \widehat\delta_{24} \, \widehat\delta_{41} \, \ra_{1234}^{\bf K_{1,2}} \propto \\
&\int \prod_i d x_i d y_i d u_i d v_i \, \Delta^2(x) \Delta^2(y) \Delta^2(u) \Delta^2(v) \, e^{ -\frac{2 N}{\lambda} \sum_i \left(x_i - \frac{K_1\lambda}{4N}\right)^2 -\frac{2 N}{\lambda} \sum_i \left(y_i + \frac{K_1\lambda}{4N}\right)^2} \\
&\times e^{ -\frac{2 N}{\lambda} \sum_i \left(u_i - \frac{K_2\lambda}{4N}\right)^2 -\frac{2 N}{\lambda} \sum_i \left(v_i + \frac{K_2\lambda}{4N}\right)^2 } \prod_{i,j} \frac{1 + e^{x_i - u_j}}{1 - e^{x_i - u_j}} \frac{1 + e^{x_i - v_j}}{1 - e^{x_i - v_j}} \frac{1 + e^{y_i - u_j}}{1 - e^{y_i - u_j}} \frac{1 + e^{y_i - v_j}}{1 - e^{y_i - v_j}}
\fe
The finite-$N$ saddle point equations of this four-matrix model are
\ieg\label{4-cover coupled saddle point equations}
\frac{4 N}{\lambda} \left(x_i-\frac{K_1\lambda}{4N}\right) = \sum_{j (\neq i)}^N \frac{2}{x_i - x_j} - \sum_{j=1}^N \frac{1}{\sinh(x_i - u_j)} - \sum_{j=1}^N \frac{1}{\sinh(x_i - v_j)}  \\
\frac{4 N}{\lambda} \left(y_i+\frac{K_1\lambda}{4N}\right) = \sum_{j(\neq i)}^N \frac{2}{y_i - y_j} - \sum_{j=1}^N \frac{1}{\sinh(y_i - u_j)} - \sum_{j=1}^N \frac{1}{\sinh(y_i - v_j)}  \\
\frac{4 N}{\lambda} \left(u_i-\frac{K_2\lambda}{4N}\right) = \sum_{j(\neq i)}^N \frac{2}{u_i - u_j} - \sum_{j=1}^N \frac{1}{\sinh(u_i - x_j)} - \sum_{j=1}^N \frac{1}{\sinh(u_i - y_j)}  \\
\frac{4 N}{\lambda} \left(v_i+\frac{K_2\lambda}{4N}\right) = \sum_{j(\neq i)}^N \frac{2}{v_i - v_j} - \sum_{j=1}^N \frac{1}{\sinh(v_i - x_j)} - \sum_{i=1}^N \frac{1}{\sinh(v_i - y_j)} 
\feg
We find that an ansatz for the bound-state eigenvalue configuration that solves \eqref{4-cover coupled saddle point equations} is
\be
(y_1 \lesssim u_1 \lesssim v_1 \lesssim x_1) < (y_2 \lesssim u_2 \lesssim v_2 \lesssim x_2) < \cdots < (y_N \lesssim u_N \lesssim v_N \lesssim x_N)
\ee
More precisely, we require that for all diagonal eigevalues $i$,
\bea\label{4-cover bound-state condition 1}
    \frac{1}{x_i - v_i} = \frac{1}{u_i - y_i} \approx \frac{K_1+K_2}{2}, \quad \frac{1}{x_i - u_i} \approx \frac{1}{v_i - y_i} = \frac{K_1-K_2}{2}
\eea 
and for all $(i,j)$ pairs where $i\neq j$,
\bea\label{4-cover bound-state condition 2}
    x_i - x_j \approx y_i - y_j \approx u_i - u_j \approx v_i - v_j \sim (i-j)\frac{\sqrt{\lambda}}{N}
\eea 
In the semiclassical regime $1\ll \lambda \ll N$, the distance between the eigenvalues satisfy
\bea
    x_i - v_i \approx v_i - u_i \approx u_i - y_i \ll x_i - x_j \approx y_i - y_j \approx u_i - u_j \approx v_i - v_j
\eea 
for all $i\neq j$. Our assumption that $K_1 - K_2 \sim K_1 + K_2 \sim O(N)$ ensures that the distances within $y$-$u$-$v$-$x$ bound state are all of order $1/N$, which is parametrically smaller than the average spacing $O(\sqrt{\lambda}/N)$ between nearest-neighbor bound states in the limit of large $\lambda$.

An immediate consequence of \eqref{4-cover bound-state condition 1} is that the order $N$ external force terms balance with the diagonal cross-terms in \eqref{4-cover coupled saddle point equations}, so that the saddle point equations reduce to
\ieg\label{4-cover still coupled saddle point equations}
\frac{4 N}{\lambda} x_i \approx \sum_{j (\neq i)}^N \frac{2}{x_i - x_j} - \sum_{j (\neq i)}^N \frac{1}{\sinh(x_i - u_j)} - \sum_{j (\neq i)}^N \frac{1}{\sinh(x_i - v_j)}  \\
\frac{4 N}{\lambda} y_i \approx \sum_{j(\neq i)}^N \frac{2}{y_i - y_j} - \sum_{j (\neq i)}^N \frac{1}{\sinh(y_i - u_j)} - \sum_{j (\neq i)}^N \frac{1}{\sinh(y_i - v_j)}  \\
\frac{4 N}{\lambda} u_i \approx \sum_{j(\neq i)}^N \frac{2}{u_i - u_j} - \sum_{j (\neq i)}^N \frac{1}{\sinh(u_i - x_j)} - \sum_{j (\neq i)}^N \frac{1}{\sinh(u_i - y_j)} \\
\frac{4 N}{\lambda} v_i \approx \sum_{j(\neq i)}^N \frac{2}{v_i - v_j} - \sum_{j (\neq i)}^N \frac{1}{\sinh(v_i - x_j)} - \sum_{j (\neq i)}^N \frac{1}{\sinh(v_i - y_j)} .
\feg
In the limit of large $\lambda$, the off-diagonal cross-terms in \eqref{4-cover still coupled saddle point equations} are negligible beyond an $O(1)$ neighborhood of each eigenvalue, and, within the $O(1)$ neighborhood, they cancel pairwise. As a result, these terms can be dropped and \eqref{4-cover still coupled saddle point equations} simplifies to
\ieg\label{4-cover decoupled saddle point equations}
\frac{4 N}{\lambda} x_i \approx \sum_{j (\neq i)}^N \frac{2}{x_i - x_j}, \quad \frac{4 N}{\lambda} y_i \approx \sum_{j(\neq i)}^N \frac{2}{y_i - y_j} \\
\frac{4 N}{\lambda} u_i \approx \sum_{j(\neq i)}^N \frac{2}{u_i - u_j}, \quad \frac{4 N}{\lambda} v_i \approx \sum_{j(\neq i)}^N \frac{2}{v_i - v_j} 
\feg
as in the $2$-cover case. The solutions are Wigner semicircles
\be\label{4-cover overlapping Wigner} 
    \rho(x) = \frac{2}{\lambda \pi} \sqrt{\lambda - x^2}, \quad \rho(y) = \frac{2}{\lambda \pi} \sqrt{\lambda - y^2}, \quad 
    \rho(u) = \frac{2}{\lambda \pi} \sqrt{\lambda - u^2}, \quad
    \rho(v) = \frac{2}{\lambda \pi} \sqrt{\lambda - v^2}
\ee
This provides an a posterori justification for our ansatz \eqref{4-cover bound-state condition 2} of an order $\sqrt{\lambda}/N$ spacing between the neighboring bound-states. To leading order, the bound states $y$-$u$-$v$-$x$ of eigenvalues result in four identified Wigner distributions, recovering the result of the 4-1 map of the four-cover BW$_4$ in Section \ref{fourregions}. In particular, see the discussion around \eqref{four to one transform}.

An alternative way to see that the bound-state solution \eqref{4-cover overlapping Wigner} reproduces the branch cuts
\ie 
    \relax [-a^2 e_1^{-1},-a^2 e_2^{-1}]\cup[-e_2,-e_1]\cup[e_1,e_2]\cup[a^2 e_2^{-1},a^2 e_1^{-1}]
\fe 
and the simple poles at $0$, $\pm a$, and $\infty$ of the BW$_4$ $z$-plane is via
\bea\label{4 branches of the 1-4 map}
    & z(x) = \frac{x}{4}-\frac{\sqrt{x^2+4 c^2}}{4}- \frac{1}{2}\sqrt{\frac{1}{2}\left(8a^2+2c^2+x^2-x\sqrt{x^2+4c^2}\right)} \\ 
    & z(y) = \frac{y}{4}+\frac{\sqrt{y^2+4c^2}}{4}-\frac{1}{2}\sqrt{\frac{1}{2}\left(8a^2+2c^2+y^2+y\sqrt{y^2+4c^2}\right)} \\
    & z(u) = \frac{u}{4}-\frac{\sqrt{u^2+4c^2}}{4}+\frac{1}{2}\sqrt{\frac{1}{2}\left(8a^2+2c^2+u^2-u\sqrt{u^2+4c^2}\right)} \\
    & z(v) = \frac{v}{4}+\frac{\sqrt{v^2+4c^2}}{4}+\frac{1}{2}\sqrt{\frac{1}{2}\left(8a^2+2c^2+v^2+v\sqrt{v^2+4c^2}\right)}
\eea 
where we have defined 
\ie 
    c^2 \equiv \frac{(a^2-e_2^2)(a^2-e_1^2)}{e_1e_2}
\fe
Using \eqref{4-cover WH lambda in terms of e2, e1, and a}, it holds that
\begin{itemize}
    \item The branch cut $[-\sqrt{\lambda},\sqrt{\lambda}]$ and the simple pole at $\infty$ in the $x$-plane map to the branch cut $[-a^2 e_1^{-1},-a^2 e_2^{-1}]$ and the simple pole at $-a$ in the $z$-plane.
    \item The branch cut $[-\sqrt{\lambda},\sqrt{\lambda}]$ and the simple pole at $\infty$ in the $y$-plane map to the branch cut $[-e_2,-e_1]$ and the simple pole at $0$ in the $z$-plane.
    \item The branch cut $[-\sqrt{\lambda},\sqrt{\lambda}]$ and the simple pole at $\infty$ in the $u$-plane map to the branch cut $[e_1,e_2]$ and the simple pole at $a$ in the $z$-plane.
    \item The branch cut $[-\sqrt{\lambda},\sqrt{\lambda}]$ and the simple pole at $\infty$ in the $v$-plane map to the branch cut $[a^2 e_2^{-1},a^2 e_1^{-1}]$ and the simple pole at $\infty$ in the $z$-plane.
\end{itemize}
Moreover, the bound-state eigenvalue density \eqref{4-cover overlapping Wigner} in the $z$-variable \eqref{4 branches of the 1-4 map} is
\ie 
    \rho(z) = \frac{2}{\lambda \pi} \sqrt{\frac{(e_2^2-z^2)(z^2-e_1^2)(z^2-a^4 e_1^{-2})(z^2-a^4 e_2^{-2})}{z^2(z-a)^2(z+a)^2}}
\fe 
which can be obtained by taking the discontinuity (see \eqref{splittingres}) of the following resolvent
\begin{equation}\label{4-cover resolvent in z-variable}
    \begin{split}
        \omega(z)=&\frac{2}{\lambda}\biggl(z-\frac{a^2}{z}-\frac{(a^2-e_2^2)(a^2-e_1^2)}{2 e_1 e_2}\left(\frac{1}{z+a}+\frac{1}{z-a}\right)\\
        &\hphantom{\frac{\mu}{2}\biggl(}
        -\frac{\sqrt{(z^2-e_1^2)(z^2-e_2^2)(z^2-a^4/e_2^2)(z^2-a^4/e_1^2)}}{z(z-a)(z+a)}\biggl)
    \end{split}
\end{equation}
across its branch cuts.\ \eqref{4-cover resolvent in z-variable} is the resolvent of the four-cover Gaussian–Penner matrix model \eqref{eqn: 4-cover GP MM}, which is discussed in more detail in Appendix~\ref{appendix: Gaussian-Penner model}.

Applying the relation \eqref{matchingrelation} to the four-cover Gaussian–Penner resolvent \eqref{4-cover resolvent in z-variable} yields precisely the harmonic functions $h_{1,2}$ of BW$_4$. However, the four-cover Gaussian–Penner model admits more general solutions than \eqref{4-cover resolvent in z-variable}. To select the specific solution \eqref{4-cover resolvent in z-variable}, we must impose the regularity conditions of the harmonic functions\footnote{There is no obstruction to considering a three-cover Gaussian–Penner model and solving for its resolvent, but the obstruction to constructing a three-cover BW$_3$ arises from the regularity conditions of the supergravity solution.}. In particular, $\partial_z h_1$ and $\partial_z h_2$ must share common zeroes in the interior of the Riemann surface $\Sigma$. 

As in BW$_2$, we can regard the four-cover Gaussian–Penner matrix model as an effective matrix model description of BW$_4$. By contrast, the four-matrix model \eqref{4-cover MM eigenvalue basis}, together with \eqref{4 branches of the 1-4 map}, constitutes our proposal for how BW$_4$ emerges from four originally decoupled Gaussian matrix models via a cyclic insertion of delta operators.

\section{Free energy and on-shell action} \label{sec: free energy}

In this section, we compute the free energy of the delta operator $\la \widehat\delta_{12} \ra_{12}^{\bf K}$ in the matrix model background \eqref{2-cover WH matrix model eigenvalue basis}. On the bulk side, we find that the leading on-shell action of a minimal bulk source that would produce the expected conical singularity cancels precisely with the contribution from the conical singularity to the Einstein-Hilbert action. We model the subleading term in the effective action of the source as a Dvali-Gabadadze-Porrati (DGP) term \cite{Dvali:2000hr}, i.e. induced gravity on the bulk source, and, while the multiplicative constants cannot be determined, we observe that the remaining on-shell action has a parametric scaling that is consistent with that of $\la \widehat\delta_{12} \ra_{12}^{\bf K}$. 

Let us compute the free energy $F_{\delta}$ of the delta operator
\be\label{2-cover delta operator eigenvalue basis}
\la \, \widehat\delta_{12} \, \ra_{12}^{\bf K} = \frac{1}{Z_1 Z_2} \int \prod_i d x_i d y_i \, \Delta^2(x) \Delta^2(y) \, e^{ - \frac{2N}{\lambda} \sum_{i} x_i^2 + K \sum_i x_i - \frac{2N}{\lambda} \sum_{i} y_i^2 - K \sum_i y_i } \prod_{i,j} \frac{1+ e^{x_i-y_j}}{1- e^{x_i-y_j}}
\ee
$F_{\delta}$ is obtained by evaluating \eqref{2-cover delta operator eigenvalue basis} on the bound-state configuration \eqref{2-cover bound-state} with the eigenvalue densities \eqref{solutions to saddlepequationsdecoupled}. Note that, according to our normalization factor $1/Z_1 Z_2$ in \eqref{2-cover delta operator eigenvalue basis}, $F_{\delta}$ is computed relative to the free energy of two decoupled $\AdS_5\times S^5$. As we will show,
\ie\label{MM delta operator free energy} 
   \la \, \widehat\delta_{12} \, \ra_{12}^{\bf K} \approx e^{- F_{\delta}}, \qquad F_{\delta} = - \frac{80}{3 \pi^2} \frac{N^2}{\sqrt{\lambda}} 
\fe 
to leading order in large $N$ and $\lambda$. The negativity of this free energy reflects the attractive nature of the delta operator. Our model for the bulk source will provide an example for how the $N^2/\sqrt{\lambda}$ dependence could arise.

Let us now derive \eqref{MM delta operator free energy}. Note that the eigenvalue densities \eqref{solutions to saddlepequationsdecoupled} are symmetric around $x=0$ and $y=0$. Thus,
\ie
     K \sum_i x_i \approx K N \int_{-\sqrt{\lambda}}^{\sqrt{\lambda}} d x \, \rho(x) x = 0 , \quad  - K \sum_i y_i \approx - K N \int_{-\sqrt{\lambda}}^{\sqrt{\lambda}} d y \, \rho(y) y = 0 
\fe
Furthermore, when evaluated on \eqref{solutions to saddlepequationsdecoupled},
\ie 
    \Delta^2(x) \Delta^2(y) \,  e^{ - \frac{2N}{\lambda} \sum_{i} x_i^2 - \frac{2N}{\lambda} \sum_{i} y_i^2 }
\fe
cancels with the normalization factor $1/Z_1 Z_2$ to leading order in large $N$ and $\lambda$. Consequently, the leading contribution comes from the cross-term in \eqref{2-cover delta operator eigenvalue basis}:
\ie\label{exact delta operator free energy} 
    \la \, \widehat\delta_{12} \, \ra_{12}^{\bf K} \approx \exp(N^2 \int_{-\sqrt{\lambda}}^{\sqrt{\lambda}} d x\, \rho(x) \dashint_{-\sqrt{\lambda}}^{\sqrt{\lambda}} d y\, \rho(y) \log(\frac{1+ e^{x-y}}{1- e^{x-y}}) )
\fe 
Here, the inner integral is understood in the principal value (PV) sense
\ie\label{definition of PV-regularized integral} 
    \dashint_{-\sqrt{\lambda}}^{\sqrt{\lambda}} d y = \lim_{\epsilon \to 0} \left( \int_{-\sqrt{\lambda}}^{x-\epsilon} d y + \int_{x+\epsilon}^{\sqrt{\lambda}} d y \right)
\fe 
as microscopically, the separation between $x_i$ and $y_j$ is approximately $1/K$ when they form a bound pair, and is typically of order $(i - j) \sqrt{\lambda}/N$ otherwise. The $N^2/\sqrt{\lambda}$ scaling of the free energy can be argued via a scaling analysis similar to that used for solving the coupled saddle point equations \eqref{saddlepequationsmixed} of our two-matrix model. The prefactor $80/3\pi^2$ can be obtained by numerically evaluating \eqref{exact delta operator free energy}.

To perform the scaling analysis, it is convenient to separate the diagonal and off-diagonal contributions in the finite-$N$ cross-term 
\ie 
    \prod_{i,j} \frac{1+ e^{x_i-y_j}}{1- e^{x_i-y_j}} = \left(\prod_i \frac{1+ e^{x_i-y_i}}{1- e^{x_i-y_i}}\right) \left(\prod_i \prod_{j (\neq i)} \frac{1+ e^{x_i-y_j}}{1- e^{x_i-y_j}}\right)
\fe
Using the fact that $K \sim O(N)$, the diagonal contribution evaluated on the bound-state configuration \eqref{2-cover bound-state} is of order $N \log N$ in the large-$N$ limit:
\ie 
    \prod_i \frac{1+ e^{x_i-y_i}}{1- e^{x_i-y_i}} \approx \left(\frac{1+ e^{\frac{1}{K}}}{1- e^{\frac{1}{K}}}\right)^N \approx (-2K)^N
\fe 
From \eqref{exact delta operator free energy}, we expect the leading contribution to be of order $N^2$ in the large-$N$ limit. Therefore, the diagonal contribution is subleading.

\begin{figure}[t]
    \centering
    \begin{subfigure}[b]{0.32\textwidth}
        \centering
        \includegraphics[width=\textwidth]{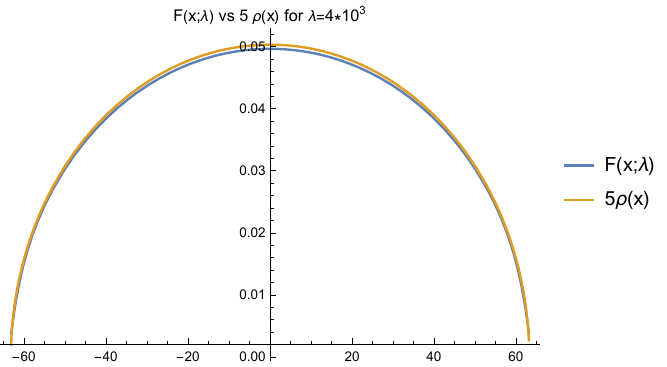}
        \caption{$\lambda = 4 \times 10^3$}
    \end{subfigure}
    \begin{subfigure}[b]{0.32\textwidth}
        \centering
        \includegraphics[width=\textwidth]{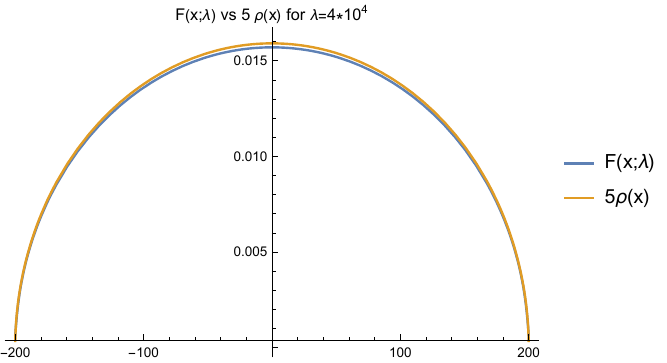}
        \caption{$\lambda = 4 \times 10^4$}
    \end{subfigure}
    \begin{subfigure}{0.32\textwidth}
        \centering
        \includegraphics[width=\linewidth]{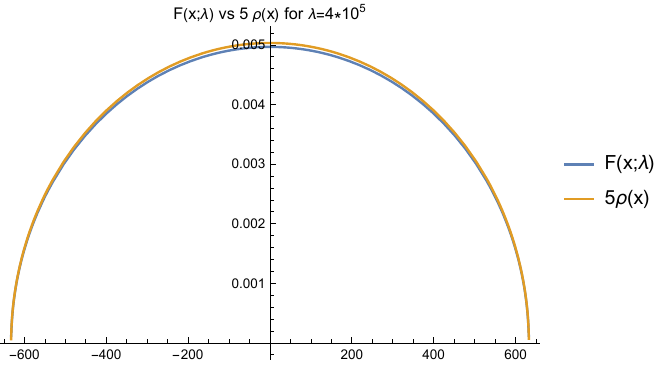}
        \caption{$\lambda = 4 \times 10^5$}
    \end{subfigure}
    
    \caption{The function $F(x;\lambda)$, defined in \eqref{definition of F(x;lambda)}, is evaluated numerically and compared with the approximation $C \rho(x)$ suggested by the scaling analysis in \eqref{approximated PV regularized integral}. The PV-regularized integral, as defined in \eqref{definition of PV-regularized integral}, is evaluated using a regulator $\epsilon=10^{-13}$. Setting $C=5$, we find good agreement $F(x;\lambda) \approx 5\rho(x)$ for large $\lambda$.}
    \label{fig: approximated PV regularized integral}
\end{figure}

To study the off-diagonal contribution, we first fix a typical $x_i$ (i.e., one not near the edges of the eigenvalue distribution) and consider the product over $y_j$ with $j\neq i$. When $x_i - y_j \gg 1$, we can approximate $(1+ e^{x_i-y_j})/(1- e^{x_i-y_j}) \approx - 1$, while for $x_i - y_j \ll 1$, one can approximate $(1+ e^{x_i-y_j})/(1- e^{x_i-y_j}) \approx - 2 / (x_i - y_j)$. The transition between these regimes occurs when $x_i - y_j \approx \alpha$, with $\alpha \sim O(1)$. This allows the approximation
\ie 
    \prod_{j (\neq i)} \frac{1+ e^{x_i-y_j}}{1- e^{x_i-y_j}} \approx \exp( N \dashint_{x_i - \alpha}^{x_i + \alpha} d y\, \rho(y) \log( - \frac{2}{x_i - y} ) )
\fe 
Since the width of the eigenvalue distribution is of order $\sqrt{\lambda}$, the $O(1)$ segment $[x_i-\alpha,x_i+\alpha]$ is approximately uniform in the large $\lambda$ limit. Hence, we can further approximate
\ie\label{approximated PV regularized integral}
    \prod_{j (\neq i)} \frac{1+ e^{x_i-y_j}}{1- e^{x_i-y_j}} \approx \exp( N \rho(x_i) \dashint_{x_i - \alpha}^{x_i + \alpha} d y\, \log( - \frac{2}{x_i - y} ) ) \sim e^{C  N\rho(x_i)}
\fe 
where $C$ is an $\alpha$-dependent $O(1)$ constant that can be obtained by evaluating the PV-regularized integral over $y$. Taking the product over all $x_i$, the off-diagonal contribution evaluated with \eqref{solutions to saddlepequationsdecoupled} becomes
\ie\label{MM free energy of delta operator with O(1)}
    \prod_i \prod_{j (\neq i)} \frac{1+ e^{x_i-y_j}}{1- e^{x_i-y_j}} \sim \prod_i e^{C N\rho(x_i)} \approx \exp(C N^2 \int_{-\sqrt{\lambda}}^{\sqrt{\lambda}} d x\,\rho(x)^2) = \exp(\frac{16 C}{3 \pi^2} \frac{N^2}{\sqrt{\lambda}})
\fe 
As a check on our scaling analysis, we numerically evaluate
\ie\label{definition of F(x;lambda)}
    F(x ; \lambda) \equiv \dashint_{-\sqrt{\lambda}}^{\sqrt{\lambda}} d y\, \rho(y) \log(\frac{1+ e^{x-y}}{1- e^{x-y}})
\fe 
and compare it with our proposed approximation $C \rho(x)$ from \eqref{approximated PV regularized integral}. We find that $F(x;\lambda) \approx 5 \rho(x)$ for large $\lambda$, as shown in Figure~\ref{fig: approximated PV regularized integral}. Consequently, \eqref{MM free energy of delta operator with O(1)} together with the numerical observation $C=5$ gives the free energy of the delta operator stated in \eqref{MM delta operator free energy}.

We now compute the on-shell contributions from a codimension-2 minimal bulk source that can source the effect of the delta function in the Ricci scalar $R_\Sigma$ of the two-cover BW$_2$ in Section \ref{tworegions}, as well as that from the Einstein-Hilbert action evaluated on $R_\Sigma$.

A minimal source that backreacts and creates a conical deficit angle $\Delta\phi = 2 \pi \frac{n-1}{n}$ on the ambient geometry can be modeled by a codimension-2 cosmic brane \cite{PhysRevD.23.852,Dong:2016fnf} of tension
\be
T_n = \frac{n-1}{4 n G_N}
\ee
and Euclidean action
\be
S_{\rm brane} = T_n \int d^{D-2} x \sqrt{g_{D-2}}.
\ee
We are interested in the case $n=1/2$ where $\Delta\phi = -2 \pi$ and $T_{1/2} = -\frac{1}{4 G_{10}} = -\frac{N^2}{2 \pi^4 L^8}$. It can be found, from the definitions in Section \ref{tworegions}, that the volumes of ${\rm EAdS}_2$, $S^2$, and $S^4$ are extremized at $z=z_c$, so the equations of motion are satisfied. The cosmic brane is therefore located at the conical singularity $z_c \in \Sigma$, which it is sourcing, and is wrapped on ${\rm EAdS}_2 \times S^2 \times S^4$. The introduction of the cosmic brane corresponds to the stress tensor
\be
T_{\mu \nu} = -\frac{1}{4 G_{10}} (g_8)_{\mu \nu} \, \delta^{(2)}(z_c)
\ee
required to cancel the excess delta function in the Ricci curvature. The on-shell action is
\be
S_{\rm brane} = \frac{16 \pi^4}{3 G_{10}} \left. f_1^2 f_2^2 f_4^4 \right\rvert_{z_c} =\frac{32}{3} N^2 t(t+1)
\ee
where the metric functions at $z_c$ are \eqref{eq: conical metrics BW2} and $t$ is an order $1$ number defined in \eqref{two-cover e1, e2 parameterization}. The leading on-shell action is $\sim N^2$ and independent of $\lambda$. Notice in particular that the minus sign from the tension $T_{1/2}$ has cancelled with the minus sign from the unit volume of ${\rm EAdS}_2$, giving an overall positive sign for $S_{\rm brane}$. 

We now compute the conical contribution $S_{\rm conical}$ to the Einstein-Hilbert action $S_{\rm EH} = -\frac{1}{16 \pi G_{10}} \int d^{10}x \sqrt{g} R$. We will ignore the regular parts that is expected to be two times those of $\AdS_5 \times S^5$, as we normalize our answer by the regular terms when we compare it with matrix model expectation values. Using the Ricci scalar $\sqrt{g_\Sigma}\, R_\Sigma = -4 \pi \delta^{(2)}(z_c)$ in \eqref{eq: ricci BW2}, the on-shell action is
\be
S_{\rm conical} = -\frac{16 \pi^4}{3 G_{10}} \left. f_1^2 f_2^2 f_4^4 \right\rvert_{z_c} = -\frac{32}{3} N^2 t(t+1)
\ee
which is precisely the opposite of the cosmic brane contribution. Therefore, the leading actions $\sim N^2$ cancel out: $S_{\rm brane} + S_{\rm conical} = 0$.

We now consider a subleading contribution to the on-shell action of a cosmic brane source and study whether it can accommodate the matrix model free energy $F_\delta \sim -N^2/\sqrt{\lambda}$ in \eqref{MM delta operator free energy}. We consider a model where the subleading term in the action of the cosmic brane is given by the DGP term \cite{Dvali:2000hr}, i.e. induced gravity on the bulk source:
\be
S_{\rm brane}' = S_{\rm brane} + S_{\rm DGP} = \int d^{8} x \sqrt{g_{8}} \left( T_{1/2} - \frac{1}{16 \pi G_8} R_8 \right).
\ee
Let us suppose in our analysis that $G_8^{-1} = \eta \alpha' G_{10}^{-1}$ where $\eta$ is an order 1 dimensionless parameter. The 8d Ricci scalar $R_8 = R_{{\rm EAdS}_2} + R_{S^2} + R_{S^4}$ on the worldvolume at the conical singularity is
\be
R_8 \big\rvert_{z_c} = \left. -\frac{2}{f_1^2} + \frac{2}{f_2^2} + \frac{12}{f_4^2} \, \right\rvert_{z_c} = \frac{12t^2 +12 t+2}{t(t+1)} \frac{1}{\alpha' \sqrt\lambda},
\ee
and we have the on-shell action
\be
S_{\rm DGP} = A \, \eta \frac{N^2}{\sqrt\lambda}
\ee
where $A = \frac{16(6t^2+6t + 1)}{3 \pi}$ is an order $1$ constant that can be absorbed into $\eta$. We find that, while there is an ambiguity $\eta$ in the multiplicative constant, the on-shell action $S_{\rm DGP} \sim \eta \, N^2/\sqrt\lambda$ scales as $F_\delta \sim -N^2/\sqrt\lambda$ for $\eta<0$. The fact that the leading $O(N^2)$ contributions cancel and leave a subleading $O(N^2/\sqrt\lambda)$ term suggests that the physics of the delta operator could be captured by dynamics that are intrinsic to the bulk source rather than by its tension alone.

\section{Probe loops} \label{sec: probe loops}

To further study the features of bubbling wormholes, we introduce probe observables. The simplest probe is a fundamental string, dual on the gauge theory side to a Wilson loop in the fundamental representation $W_\Box$. In this section, we present a bulk calculation of $\langle W_\Box \rangle_{\text{BW}}$ in bubbling wormholes and suggest their duals in the multi-matrix models.

It is also possible to compute $\langle W_R \rangle_{\text{BW}}$ for higher but non-backreacting representations, such as the $k$-th rank antisymmetric representation $R=A_k$ (corresponding to a probe D5-brane) and the $k$-th rank symmetric representation $R=S_k$ (corresponding to a probe D3-brane). Furthermore, one could compute correlators of probe Wilson loops or of probe local operators. Such calculations would further elucidate the properties of bubbling wormholes, but we leave this study for future work.

\subsection{Probe string in BW$_2$}

Let us introduce a fundamental string into the two-cover BW$_2$ (Section \ref{tworegions}) and search for minimal-area solutions. The corresponding on-shell action can be related to the expectation value of a fundamental Wilson loop $W_\Box$ in the two-matrix model \eqref{2-cover WH matrix model eigenvalue basis}. More precisely, in the large ’t Hooft coupling limit,
\bea 
    \langle W_\Box \rangle_{\text{BW}_2} \approx e^{- S_\text{on-shell}(z^*)}
\eea 
Here, $\langle \cdots \rangle_{\text{BW}_2}$ is shorthand for
\bea\label{2-matrix model expectation value}
    \langle \cdots \rangle_{\text{BW}_2} = \frac{\la \, \widehat\delta_{12} \cdots \, \ra_{12}^{\bf K}}{\la \, \widehat\delta_{12} \, \ra_{12}^{\bf K}} 
\eea 
$S$ is the Nambu-Goto action for the fundamental string in the two-cover wormhole, and $z^*$ is a saddle point at which the worldsheet area is minimized.

The minimal-area solutions for a fundamental string in bubbling solutions were worked out in \cite{Aguilera-Damia:2017znn}. Their results can be transferred verbatim to our case, simply by substituting our two-cover wormhole harmonic functions \eqref{twoboundarysolution} into their general expressions. In what follows, we briefly review their derivation and refer the reader to their work for more details.

We consider the worldsheet of the fundamental string with disk topology extending all along the $\AdS_2$ factor
\bea 
    d s_{\AdS_2}^2 = d \varrho^2 + \sinh^2 \varrho d \phi^2 \, ,
\eea 
of the bubbling wormhole geometry \eqref{eq: metric ansatz}. The $S^1$ boundary of this worldsheet is where the fundamental Wilson loop $W_\Box$ resides. Assuming the coordinates $(z,\bar{z})$ of the Riemann surface $\Sigma$ depend only on the worldsheet coordinate $\varrho$, the Nambu-Goto action becomes
\bea\label{Nambu-Goto for bubbling geometries}
    S = \frac{1}{2\pi \alpha'} \int d \phi d \varrho \sinh \varrho \, e^\phi f_1^2 \sqrt{1 + \frac{4 \rho^2}{f_1^2} \abs{\dv{z}{\varrho}}^2 } + \frac{1}{2 \pi \alpha'} \int d \phi d \varrho \sinh \varrho \,  b_1
\eea
It was shown in \cite{Aguilera-Damia:2017znn} that $z(\varrho)=z^*$ (constant) is a solution to the equation of motion if
\bea\label{minimal area condition}
    \partial_z (e^{\phi} f_1^2) \big|_{z=z^*} = \partial_z b_1 \big|_{z=z^*} = 0
\eea 
For these solutions, the on-shell action is
\bea\label{minimal area on-shell action}
S_\text{on-shell}(z^*) = - \frac{1}{\alpha'} \left( e^{\phi} f_1^2 + b_1 \right) \big|_{z=z^*} \, .
\eea
For the bubbling wormhole solutions, $b_1=0$ identically. In addition, in our normalization of the harmonic functions, $e^\phi = 1$. Thus, the condition \eqref{minimal area condition} and the on-shell action \eqref{minimal area on-shell action} simplify to
\bea 
    \partial_z(f_1^2)\big|_{z=z^*} = 0, \quad S_\text{on-shell}(z^*) = - \frac{1}{\alpha'}  f_1^2 \big|_{z=z^*}
\eea
For the two-cover BW$_2$, the metric component $f_1^2$ is stationary over the branch cuts
\bea 
    z^* = [ - e_2 , - e_1 ] \cup [ e_1, e_2 ]
\eea
as well as at the conical singularity
\bea 
    z^* = - i \sqrt{e_1 e_2}.
\eea

Let us first consider the worldsheet action evaluted on the branch cuts. Over the branch cuts $z^* = [ - e_2 , - e_1 ] \cup [ e_1, e_2 ]$, the on-shell action is
\bea 
    S_\text{on-shell}(z^*) = - ( e_2 - e_1 ) = - \sqrt{\lambda},
\eea
using the result \eqref{2-cover WH lambda = e2 - e1}. Note that this is the same as the on-shell action of a fundamental string in $\AdS_5 \times S^5$. This is expected: passing to the double cover of $\AdS_5 \times S^5$ does not change the minimal worldsheet area of a string ending on the common $S^1 = S_1^4 \cap S_2^4$ of the two asymptotic boundaries $S_1^4$ and $S_2^4$.

On the boundary, it is natural to identify these probe strings with the fundamental Wilson loops $W_\Box^{(1)} = \Tr(e^{M_1})$ and $W_\Box^{(2)} = \Tr(e^{M_2})$ defined using the gauge fields intrinsic to SYM$_1$ and SYM$_2$, respectively. For example, we have
\bea
    \langle \Tr(e^{M_1}) \rangle_{\text{BW}_2} = \left(\frac{2}{\lambda \pi}\right)^2 \int_{-\sqrt{\lambda}}^{\sqrt{\lambda}} d x d y\, e^x \sqrt{\lambda - x^2} \sqrt{\lambda - y^2} = \frac{2}{\sqrt{\lambda}} I_1(\sqrt{\lambda}).
\eea 
so that $\la W_\Box^{(1)} \ra_{\text{BW}_2} = \la W_\Box^{(2)} \ra_{\text{BW}_2} \approx e^{\sqrt{\lambda}}$, to leading order in large $N$ and $\lambda$, for BW$_2$.

The difference between BW$_2$ and the $\AdS_5 \times S^5$ geometry can only be revealed by observables that probe the double-cover structure. As mentioned, there is another saddle point for the Nambu-Goto action \eqref{Nambu-Goto for bubbling geometries}, localized at the conical singularity $z^* =  - i \sqrt{e_1 e_2}$. At this location, the on-shell action is
\bea 
S_\text{on-shell}\left(z^*\right) = - \frac{(e_2 + e_1)^2}{e_2 - e_1} = - \sqrt{\lambda} (1+t) \, ,
\eea 
where, with the relation $e_2-e_1=\sqrt{\lambda}$, we parameterize  $e_1$ and $e_2$ by an $O(1)$ parameter $t>0$ as follows:
\be\label{two-cover e1, e2 parameterization}
e_1 = \frac{\sqrt{\lambda}}{2} ( \sqrt{1 + t}-1), \quad e_2 = \frac{\sqrt{\lambda}}{2} ( \sqrt{1 + t}+1).
\ee
The expectation value of a string at the conical singularity dominates over those on the branch cuts. 

A possibility for the boundary dual to a probe string at the conical singularity is a Wilson loop $W_\Box^{(12)}$ constructed from a ``diagonal'' combination of the gauge connections. Upon localization, this becomes $\Tr_\Box(e^{M})$ where $M = M_1 + t M_2$ where $t = n_2/n_1$ for integers $n_1, n_2$, and has the expectation value $\la \Tr_\Box(e^{M}) \ra_{{\rm BW}_2} \approx e^{\sqrt{\lambda} (1+t)}$ to leading order in large $N$ and $\lambda$.
This matches the on-shell string action when the parameter $t$ in the operator agrees with the geometric parameter in \eqref{two-cover e1, e2 parameterization}. We speculate that this correlation reflects an ambiguity in defining the diagonal direction when lifting to the covering geometry.

\subsection{Probe string in BW$_4$}

For the four-cover BW$_4$, the metric component $f_1^2$ is stationary over the branch cuts
\bea\label{4-cover branch cuts} 
    z^* = [-a^2 e_1^{-1},-a^2 e_2^{-1}]\cup[-e_2,-e_1]\cup[e_1,e_2]\cup[a^2 e_2^{-1},a^2 e_1^{-1}]
\eea 
as well as at the conical singularities
\ieg
z^* = -ia \\
z^* = -\frac{i}{2}\left(\sqrt{\frac{(a^2-e_1^2)(a^2-e_2^2)}{e_1e_2}} \pm \sqrt{\frac{(a^2-e_1^2)(a^2-e_2^2)}{e_1 e_2} -4a^2}\right)
\feg
Over the branch cuts \eqref{4-cover branch cuts}, the on-shell action is
\bea 
    S_\text{on-shell}(z^*) = - \frac{(e_2-e_1)(a^2+e_1e_2)}{e_1e_2} = - \sqrt{\lambda} 
\eea
using \eqref{4-cover WH lambda in terms of e2, e1, and a}. This is the same as the on-shell action of a fundamental string in $\AdS_5 \times S^5$ as expected. Again, it is natural to identify these probe strings with the fundamental Wilson loops $W_\Box^{(i)} = \Tr(e^{M_i})$ defined using only the gauge fields intrinsic to SYM$_i$ where $i=1,2,3,4$. Their expectation values are $\la \Tr(e^{M_i}) \ra_{{\rm BW}_4} \approx e^{\sqrt\lambda}$.

Analogous to our parametrization of the two-cover parameters $e_1$ and $e_2$ by $t>0$ in \eqref{two-cover e1, e2 parameterization}, we find it useful to introduce parameters $t_1,t_2>0$ for the four-cover case. The four-cover parameters $e_1$, $e_2$, and $a$ satisfy the relation \eqref{4-cover WH lambda in terms of e2, e1, and a}:
\bea 
    \lambda = \frac{ (e_2 - e_1)^2 (a^2 + e_1 e_2)^2}{e_1^2 e_2^2}
\eea 
We parametrize them implicitly via the equations:
\bea 
    \left( \frac{a^2}{e_1} - e_1 \right) \left( \frac{a^2}{e_2} - e_2 \right) = \frac{\lambda t_1}{2}, \quad a^2 = \frac{\lambda t_2}{4}
\eea 
In this parametrization, the conical singularities are located at
\bea 
    z_1 \coloneqq - i \frac{\sqrt{\lambda t_2}}{2}, \quad z_\pm \coloneqq - i \frac{\sqrt{\lambda}}{2} \left( \sqrt{\frac{t_1}{2}} \pm \sqrt{\frac{t_1}{2}-t_2} \right)
\eea 
For a fixed $t_2>0$, the locations of $z_\pm$ change as we increase $t_1 \in (0,\infty)$. In more detail, $z_-$ starts at $z=-\sqrt{\lambda t_2}/2$ and traces out a quarter-circle in the third quadrant to reach $z=-i\sqrt{\lambda t_2}/2$ as $t_1$ increases from 0 to $2t_2$. As we further increase $t_1$ from $2t_2$ to infinity, $z_-$ moves along the imaginary axis from $z=-i\sqrt{\lambda t_2}/2$ to $z=0$. Similarly, $z_+$ starts at $z=\sqrt{\lambda t_2}/2$ and traces out a quarter-circle in the fourth quadrant to reach $z=-i\sqrt{\lambda t_2}/2$ as $t_1$ increases from 0 to $2t_2$. As we further increase $t_1$ from $2t_2$ to infinity, $z_+$ moves along the imaginary axis from $z=-i\sqrt{\lambda t_2}/2$ to $z=-i\infty$. The case $t_1=2t_2$ is special, as all three conical singularities meet at the point $z = -i\sqrt{\lambda t_2}/2$.

At the conical singularity $z^*=z_1$, the on-shell action is
\bea
    S_\text{on-shell}\left(z^*\right) =  - \frac{(a^2+e_1^2)^2(a^2+e_2^2)^2}{4a^2e_1e_2(e_2-e_1)(a^2+e_1e_2)}
    = - \sqrt{\lambda} \left( 1 + \frac{(t_1 + 2t_2)^2}{4t_2} \right),
\eea
and at $z^*=z_\pm$, the on-shell action is
\bea
    S_\text{on-shell}\left(z^* \right)
        = - \frac{(e_1+e_2)^2 (a^2-e_1 e_2)^2}{e_1 e_2 (e_2-e_1)
    (a^2+e_1 e_2)} = - \sqrt{\lambda} (1+2t_1).
\eea
The contribution from the conical singularity $z_1$ dominates over that from the conical singularities $z_\pm$, except in the special case $t_1=2t_2$ where they are equal. Moreover, the contributions from all three conical singularities dominate over that from the branch cuts.

As in BW$_2$, candidate boundary duals to the probe strings can be constructed via ``diagonal'' combinations of the gauge connections. Due to the symmetry of BW$_4$ under $z \to -z$ as written in Section \ref{fourregions}, there is a further constraint. We take $M_1$ to correspond to the simple pole at $z=0$, $M_{2,3}$ to the simple poles at $z=\pm a$, and $M_4$ to the simple pole at $z=\infty$. At $z_\pm$, we have
\be
\la \Tr(e^{M_1 + t_1 M_2 + t_1 M_3}) \ra_{\text{BW}_4} \approx e^{\sqrt{\lambda}(1 + 2 t_1)}
\ee
and at $z_1$, we have
\be
\la\Tr (e^{M_1 + \frac{1}{4 t_2}(t_1 + 2 t_2)^2 M_4}) \ra_{\text{BW}_4} \approx e^{\sqrt{\lambda} ( 1 + \frac{1}{4 t_2}(t_1 + 2 t_2)^2 )}.
\ee
We leave a more systematic study for future work.

\section*{Acknowledgements}\label{ACKNOWL}

We wish to thank Ofer Aharony, Costas Bachas, Davide Gaiotto, Jaume Gomis, Rob Myers, Carlos Nunez and Gordon Semenoff for helpful discussions. We also happily acknowledge useful discussions on related topics with the string theory groups at the University of British Columbia and Perimeter Institute, as well as with the participants of the \href{https://danninos.wixsite.com/psi-conf/psi2023}{Physics Sessions Initiative}, at the initial stages of this work. 

P.B. acknowledges financial support from the European Research Council (via the grant BHHQG-101040024), funded by the European Union.
The work of the group at ETH (J.H.L.) is supported in part by the Simons Foundation grant 994306 (Simons Collaboration on Confinement and QCD Strings) and by the NCCR SwissMAP funded by the Swiss National Science Foundation. 

Views and opinions expressed are those of the authors only and do not necessarily reflect those of the European Union or the European Research Council. Neither the European Union nor the granting authority can be held responsible for them.

\appendix

\section{The Gaussian-Penner matrix model}
\label{appendix: Gaussian-Penner model}

In this appendix, we define the Gaussian–Penner matrix model and solve for its planar resolvent. This model is a Hermitian single-matrix model with a Gaussian potential $\Tr(M^2)$ deformed by a sum of logarithmic potentials $\sum_i \Tr(\log(M-a_i)^2)$. It generalizes \cite{Tan:1991ay} the original Penner matrix model, which was studied in the context of the moduli space of punctured surfaces \cite{Penner:1986, Penner:1988cza, Harer1986}. By choosing appropriate parameters $a_i$ for the logarithmic potentials, the resulting planar resolvent maps to the harmonic functions of the bubbling wormhole via the relation \eqref{matchingrelation}. Consequently, the Gaussian–Penner matrix model provides an effective matrix model description of our bubbling wormholes.

\subsection{Two-cover Gaussian-Penner}

Consider the Gaussian-Penner matrix model corresponding to the two-cover bubbling wormhole
\be
\label{eqn: 2-cover GP MM}
Z_2=\int \mathcal{D} M\, e^{-N \Tr V_2(M)}, \quad V_2(M) = \frac{2}{\lambda} M^2-\frac{t}{2}\log M^2
\ee
The derivative of the potential is
\be
    V_2'(s)=\frac{4s}{\lambda} - \frac{t}{s}
\ee
The large-$N$ saddle point equation is
\be 
V_2'(x)=2 \slashed{\omega}_2(x),\quad \omega_2(z)=\int d s\,\frac{\rho_2(s)}{z-s}
\ee
The resolvent is solved by
\be 
\omega_2(z)=\oint_\mathcal{C}\frac{d s}{2\pi i}\, \frac{V_2'(s)/2}{z-s}\sqrt{\frac{(z^2-e_1^2)(z^2-e_2^2)}{(s^2-e_1^2)(s^2-e_2^2)}}
\ee 
where $\mathcal{C}$ is a counter-clockwise contour surrounding the branch cuts $[-e_2,-e_1]\cup[e_1,e_2]$. The result is
\be
\omega_2(z)=\frac{2 z}{\lambda}-\frac{t}{2z}-\frac{t}{2e_1e_2z}\sqrt{(z^2-e_1^2)(z^2-e_2^2)}
\ee
The first two terms come from the residue at $s=z$ and the last term comes from the residue at $s=0$. There is no residue at $s=\infty$ because the integrand scales as $s^{-2}$ there. The resolvent asymptotic conditions can be read off from
\be
\omega_2(z\to\infty)=\left(-\frac{t}{2e_1e_2}+\frac{2}{\lambda}\right)z+\frac{(e_2-e_1)^2t}{4e_1e_2z}+O(z^{-2})\stackrel{!}{=}\frac{1}{z}+O(z^{-2})
\ee
The solution is
\be 
e_1^2=\frac{\lambda}{4}\left(\sqrt{1+t}-1\right)^2,\quad e_2^2=\frac{\lambda}{4}\left(\sqrt{1+t}+1\right)^2
\ee

The final form of the resolvent is
\be
\omega_2(z)=\frac{2}{\lambda}\left(z-\frac{e_1e_2}{z}-\frac{1}{z}\sqrt{(z^2-e_1^2)(z^2-e_2^2)}\right)
\ee
with
\be 
    e_2-e_1=\sqrt{\lambda}, \quad e_1e_2=\frac{\lambda t}{4}
\ee

\subsection{Four-cover Gaussian-Penner}

Consider the Gaussian-Penner matrix model corresponding to the four-cover bubbling wormhole
\be 
\label{eqn: 4-cover GP MM}
    Z_4=\int \mathcal{D} M\,e^{-N\Tr V_4(M)}, \quad V_4(M) = \frac{2}{\lambda} M^2-\frac{t_1}{2}\log(M-a)^2-\frac{t_1}{2}\log(M+a)^2-\frac{t_2}{2}\log M^2
\ee
The derivative of the potential is
\be 
    V_4'(s)=\frac{4s}{\lambda} - \frac{t_1}{s-a} - \frac{t_1}{s+a} - \frac{t_2}{s}   
\ee 
The large-$N$ saddle point equation is
\be 
    V_4'(x)=2 \slashed{\omega}_4(x),\quad \omega_4(z)=\int d s\,\frac{\rho_4(s)}{z-s}
\ee 
The resolvent is solved by
\be
    \omega_4(z)=\oint_\mathcal{C}\frac{d s}{2\pi i}\, \frac{V_4'(s)/2}{z-s}\sqrt{\frac{(z^2-e_1^2)(z^2-e_2^2)(z^2-e_3^2)(z^2-e_4^2)}{(s^2-e_1^2)(s^2-e_2^2)(s^2-e_3^2)(s^2-e_4^2)}}
\ee  
where $\mathcal{C}$ is a counter-clockwise contour surrounding the branch cuts
\be
    [-e_4,-e_3]\cup[-e_2,-e_1]\cup[e_1,e_2]\cup[e_3,e_4]
\ee 
The result is
\begin{equation}
    \begin{split}
        \omega_4(z)=&\frac{2 z}{\lambda}-\frac{t_1}{2(z-a)}-\frac{t_1}{2(z+a)}-\frac{t_2}{2z}+\frac{t_2}{2e_1e_2e_3e_4}\frac{\sqrt{(z^2-e_1^2)(z^2-e_2^2)(z^2-e_3^2)(z^2-e_4^2)}}{z}\\
        &-\frac{t_1}{2\sqrt{(e_4^2-a^2)(e_3^2-a^2)(a^2-e_2^2)(a^2-e_1^2)}}\frac{\sqrt{(z^2-e_1^2)(z^2-e_2^2)(z^2-e_3^2)(z^2-e_4^2)}}{z-a}\\
        &-\frac{t_1}{2\sqrt{(e_4^2-a^2)(e_3^2-a^2)(a^2-e_2^2)(a^2-e_1^2)}}\frac{\sqrt{(z^2-e_1^2)(z^2-e_2^2)(z^2-e_3^2)(z^2-e_4^2)}}{z+a}
    \end{split}
\end{equation} 
Expanding the resolvent at $z=\infty$, we obtain an $O(z^3)$ term
\be\label{4-cover resolvent condition 1} 
    \omega_4(z\to\infty)=\left(\frac{t_2}{2e_1e_2e_3e_4}-\frac{t_1}{\sqrt{(e_4^2-a^2)(e_3^2-a^2)(a^2-e_2^2)(a^2-e_1^2)}}\right)z^3+O(z)
\ee 
Setting this term to 0 simplifies the resolvent 
\begin{equation}
    \begin{split}
        \omega_4(z)=&\frac{2 z}{\lambda}-\frac{t_1}{2(z-a)}-\frac{t_1}{2(z+a)}-\frac{t_2}{2z}+\frac{t_2}{2e_1e_2e_3e_4}\frac{\sqrt{(z^2-e_1^2)(z^2-e_2^2)(z^2-e_3^2)(z^2-e_4^2)}}{z}\\
        &-\frac{t_2}{4e_1e_2e_3e_4}\frac{\sqrt{(z^2-e_1^2)(z^2-e_2^2)(z^2-e_3^2)(z^2-e_4^2)}}{z-a}\\
        &-\frac{t_2}{4e_1e_2e_3e_4}\frac{\sqrt{(z^2-e_1^2)(z^2-e_2^2)(z^2-e_3^2)(z^2-e_4^2)}}{z+a}
    \end{split}
\end{equation}
Expanding the simplified resolvent at $z=\infty$, we obtain an $O(z)$ term
\be\label{4-cover resolvent condition 2} 
    \omega_4(z\to\infty)=\frac{\frac{4}{\lambda} e_1e_2e_3e_4 - a^2t_2}{2e_1e_2e_3e_4}z+O(z^{-1})
\ee 
Setting this term to 0 further simplifies the resolvent
\begin{equation}
    \begin{split}
        \omega_4(z)=&\frac{2}{\lambda}\biggl(z-\frac{e_1e_2e_3e_4}{a^2z}-\frac{\sqrt{(e_4^2-a^2)(e_3^2-a^2)(a^2-e_2^2)(a^2-e_1^2)}}{2a^2}\left(\frac{1}{z+a}+\frac{1}{z-a}\right)\\
        &\hphantom{\frac{\mu}{2}\biggl(}
        -\frac{\sqrt{(z^2-e_1^2)(z^2-e_2^2)(z^2-e_3^2)(z^2-e_4^2)}}{z(z-a)(z+a)}\biggl)
    \end{split}
\end{equation}
This is the general solution to the large-$N$ saddle point equation before fixing the normalization of the density $\rho_4(s)$.

To connect with the harmonic functions $h_{1,2}$ of the four-cover bubbling wormhole, we impose the conditions
\be\label{4-cover regularity condition}
    e_3=\frac{a^2}{e_2},\quad e_4=\frac{a^2}{e_1}
\ee 
which reduce the resolvent to
\begin{equation}\label{4-cover Gaussian-Penner resolvent}
    \begin{split}
        \omega_4(z)=&\frac{2}{\lambda}\biggl(z-\frac{a^2}{z}-\frac{(a^2-e_2^2)(a^2-e_1^2)}{2 e_1 e_2}\left(\frac{1}{z+a}+\frac{1}{z-a}\right)\\
        &\hphantom{\frac{\mu}{2}\biggl(}
        -\frac{\sqrt{(z^2-e_1^2)(z^2-e_2^2)(z^2-a^4/e_2^2)(z^2-a^4/e_1^2)}}{z(z-a)(z+a)}\biggl)
    \end{split}
\end{equation}
Conditions \eqref{4-cover regularity condition} are necessary so that the harmonic functions $h_{1,2}$, which can be read off from \eqref{4-cover Gaussian-Penner resolvent} using the relation \eqref{matchingrelation}, satisfy the regularity conditions of the supergravity solution.

Normalizing the density to unity, i.e., requiring $\omega_4(z\to\infty)=z^{-1}+O(z^{-2})$, gives the condition
\be
    \lambda = \frac{(e_2 - e_1)^2 (a^2 + e_1 e_2)^2}{e_1^2 e_2^2}
\ee
Together with the additional resolvent conditions \eqref{4-cover resolvent condition 1} and \eqref{4-cover resolvent condition 2}:
\bea 
    \left( \frac{a^2}{e_1} - e_1 \right) \left( \frac{a^2}{e_2} - e_2 \right) = \frac{2 a^2 t_1}{t_2}, \quad a^2 = \frac{\lambda t_2}{4}
\eea 
we obtain two equations for the two unknowns $e_1$ and $e_2$, as well as a relation among the matrix model parameters $\lambda$, $t_1$, $t_2$, and $a$.

\bibliography{references}

@article{DHoker:2007mci,
    author = "D'Hoker, Eric and Estes, John and Gutperle, Michael",
    title = "{Gravity duals of half-BPS Wilson loops}",
    eprint = "0705.1004",
    archivePrefix = "arXiv",
    primaryClass = "hep-th",
    reportNumber = "UCLA-07-TEP-11",
    doi = "10.1088/1126-6708/2007/06/063",
    journal = "JHEP",
    volume = "06",
    pages = "063",
    year = "2007"
}

@article{Okuda:2007kh,
    author = "Okuda, Takuya",
    title = "{A Prediction for bubbling geometries}",
    eprint = "0708.3393",
    archivePrefix = "arXiv",
    primaryClass = "hep-th",
    reportNumber = "NSF-KITP-07-164",
    doi = "10.1088/1126-6708/2008/01/003",
    journal = "JHEP",
    volume = "01",
    pages = "003",
    year = "2008"
}

@article{Okuda:2008px,
    author = "Okuda, Takuya and Trancanelli, Diego",
    title = "{Spectral curves, emergent geometry, and bubbling solutions for Wilson loops}",
    eprint = "0806.4191",
    archivePrefix = "arXiv",
    primaryClass = "hep-th",
    reportNumber = "NSF-KITP-08-89",
    doi = "10.1088/1126-6708/2008/09/050",
    journal = "JHEP",
    volume = "09",
    pages = "050",
    year = "2008"
}

@article{Yamaguchi:2006te,
    author = "Yamaguchi, Satoshi",
    title = "{Bubbling geometries for half BPS Wilson lines}",
    eprint = "hep-th/0601089",
    archivePrefix = "arXiv",
    doi = "10.1142/S0217751X07035070",
    journal = "Int. J. Mod. Phys. A",
    volume = "22",
    pages = "1353--1374",
    year = "2007"
}

@article{Faraggi:2011bb,
    author = "Faraggi, Alberto and Pando Zayas, Leopoldo A.",
    title = "{The Spectrum of Excitations of Holographic Wilson Loops}",
    eprint = "1101.5145",
    archivePrefix = "arXiv",
    primaryClass = "hep-th",
    reportNumber = "MCTP-11-02",
    doi = "10.1007/JHEP05(2011)018",
    journal = "JHEP",
    volume = "05",
    pages = "018",
    year = "2011"
}

@article{Benichou:2011aa,
    author = "Benichou, Raphael and Estes, John",
    title = "{Geometry of Open Strings Ending on Backreacting D3-Branes}",
    eprint = "1112.3035",
    archivePrefix = "arXiv",
    primaryClass = "hep-th",
    doi = "10.1007/JHEP03(2012)025",
    journal = "JHEP",
    volume = "03",
    pages = "025",
    year = "2012"
}

@article{Aguilera-Damia:2017znn,
    author = "Aguilera-Damia, Jerem{\'\i}as and Correa, Diego H. and Fucito, Francesco and Giraldo-Rivera, Victor I. and Morales, Jose F. and Pando Zayas, Leopoldo A.",
    title = "{Strings in Bubbling Geometries and Dual Wilson Loop Correlators}",
    eprint = "1709.03569",
    archivePrefix = "arXiv",
    primaryClass = "hep-th",
    reportNumber = "MCTP-17-16",
    doi = "10.1007/JHEP12(2017)109",
    journal = "JHEP",
    volume = "12",
    pages = "109",
    year = "2017"
}

@article{DHoker:2007hhe,
    author = "D'Hoker, Eric and Estes, John and Gutperle, Michael",
    title = "{Exact half-BPS Type IIB interface solutions. II. Flux solutions and multi-Janus}",
    eprint = "0705.0024",
    archivePrefix = "arXiv",
    primaryClass = "hep-th",
    reportNumber = "UCLA-07-TEP-10",
    doi = "10.1088/1126-6708/2007/06/022",
    journal = "JHEP",
    volume = "06",
    pages = "022",
    year = "2007"
}

@article{DHoker:2007zhm,
    author = "D'Hoker, Eric and Estes, John and Gutperle, Michael",
    title = "{Exact half-BPS Type IIB interface solutions. I. Local solution and supersymmetric Janus}",
    eprint = "0705.0022",
    archivePrefix = "arXiv",
    primaryClass = "hep-th",
    reportNumber = "UCLA-07-TEP-09",
    doi = "10.1088/1126-6708/2007/06/021",
    journal = "JHEP",
    volume = "06",
    pages = "021",
    year = "2007"
}

@article{Erickson:2000af,
    author = "Erickson, J. K. and Semenoff, G. W. and Zarembo, K.",
    title = "{Wilson loops in N=4 supersymmetric Yang-Mills theory}",
    eprint = "hep-th/0003055",
    archivePrefix = "arXiv",
    reportNumber = "ITEP-TH-13-00",
    doi = "10.1016/S0550-3213(00)00300-X",
    journal = "Nucl. Phys. B",
    volume = "582",
    pages = "155--175",
    year = "2000"
}

@article{Drukker:2000rr,
    author = "Drukker, Nadav and Gross, David J.",
    title = "{An Exact prediction of N=4 SUSYM theory for string theory}",
    eprint = "hep-th/0010274",
    archivePrefix = "arXiv",
    reportNumber = "USC-00-05, CITUSC-00-057, NSF-ITP-00-119",
    doi = "10.1063/1.1372177",
    journal = "J. Math. Phys.",
    volume = "42",
    pages = "2896--2914",
    year = "2001"
}

@article{Maldacena:1998im,
    author = "Maldacena, Juan Martin",
    title = "{Wilson loops in large N field theories}",
    eprint = "hep-th/9803002",
    archivePrefix = "arXiv",
    reportNumber = "HUTP-98-A014",
    doi = "10.1103/PhysRevLett.80.4859",
    journal = "Phys. Rev. Lett.",
    volume = "80",
    pages = "4859--4862",
    year = "1998"
}

@article{Rey:1998ik,
    author = "Rey, Soo-Jong and Yee, Jung-Tay",
    title = "{Macroscopic strings as heavy quarks in large N gauge theory and anti-de Sitter supergravity}",
    eprint = "hep-th/9803001",
    archivePrefix = "arXiv",
    reportNumber = "SNUTP-98-016",
    doi = "10.1007/s100520100799",
    journal = "Eur. Phys. J. C",
    volume = "22",
    pages = "379--394",
    year = "2001"
}

@article{Drukker:1999zq,
    author = "Drukker, Nadav and Gross, David J. and Ooguri, Hirosi",
    title = "{Wilson loops and minimal surfaces}",
    eprint = "hep-th/9904191",
    archivePrefix = "arXiv",
    reportNumber = "UCB-PTH-99-11, LBNL-43079, NSF-ITP-99-22, LBL-43079",
    doi = "10.1103/PhysRevD.60.125006",
    journal = "Phys. Rev. D",
    volume = "60",
    pages = "125006",
    year = "1999"
}

@article{Semenoff:2001xp,
    author = "Semenoff, Gordon W. and Zarembo, K.",
    title = "{More exact predictions of SUSYM for string theory}",
    eprint = "hep-th/0106015",
    archivePrefix = "arXiv",
    reportNumber = "ITEP-TH-28-01",
    doi = "10.1016/S0550-3213(01)00455-2",
    journal = "Nucl. Phys. B",
    volume = "616",
    pages = "34--46",
    year = "2001"
}

@article{Zarembo:2002an,
    author = "Zarembo, K.",
    title = "{Supersymmetric Wilson loops}",
    eprint = "hep-th/0205160",
    archivePrefix = "arXiv",
    doi = "10.1016/S0550-3213(02)00693-4",
    journal = "Nucl. Phys. B",
    volume = "643",
    pages = "157--171",
    year = "2002"
}

@article{Pestun:2007rz,
    author = "Pestun, Vasily",
    title = "{Localization of gauge theory on a four-sphere and supersymmetric Wilson loops}",
    eprint = "0712.2824",
    archivePrefix = "arXiv",
    primaryClass = "hep-th",
    reportNumber = "ITEP-TH-41-07, PUTP-2248",
    doi = "10.1007/s00220-012-1485-0",
    journal = "Commun. Math. Phys.",
    volume = "313",
    pages = "71--129",
    year = "2012"
}

@article{Zarembo:2016bbk,
    author = "Zarembo, Konstantin",
    title = "{Localization and AdS/CFT Correspondence}",
    eprint = "1608.02963",
    archivePrefix = "arXiv",
    primaryClass = "hep-th",
    reportNumber = "NORDITA-2016-30, UUITP-08-16",
    doi = "10.1088/1751-8121/aa585b",
    journal = "J. Phys. A",
    volume = "50",
    number = "44",
    pages = "443011",
    year = "2017"
}

@article{Drukker:2005kx,
    author = "Drukker, Nadav and Fiol, Bartomeu",
    title = "{All-genus calculation of Wilson loops using D-branes}",
    eprint = "hep-th/0501109",
    archivePrefix = "arXiv",
    reportNumber = "ITFA-2005-01",
    doi = "10.1088/1126-6708/2005/02/010",
    journal = "JHEP",
    volume = "02",
    pages = "010",
    year = "2005"
}

@article{Fiol:2013hna,
    author = "Fiol, Bartomeu and Torrents, Gen{\'\i}s",
    title = "{Exact results for Wilson loops in arbitrary representations}",
    eprint = "1311.2058",
    archivePrefix = "arXiv",
    primaryClass = "hep-th",
    doi = "10.1007/JHEP01(2014)020",
    journal = "JHEP",
    volume = "01",
    pages = "020",
    year = "2014"
}

@article{Hartnoll:2006is,
    author = "Hartnoll, Sean A. and Kumar, S. Prem",
    title = "{Higher rank Wilson loops from a matrix model}",
    eprint = "hep-th/0605027",
    archivePrefix = "arXiv",
    doi = "10.1088/1126-6708/2006/08/026",
    journal = "JHEP",
    volume = "08",
    pages = "026",
    year = "2006"
}

@article{Yamaguchi:2006tq,
    author = "Yamaguchi, Satoshi",
    title = "{Wilson loops of anti-symmetric representation and D5-branes}",
    eprint = "hep-th/0603208",
    archivePrefix = "arXiv",
    doi = "10.1088/1126-6708/2006/05/037",
    journal = "JHEP",
    volume = "05",
    pages = "037",
    year = "2006"
}

@article{Gomis:2006sb,
    author = "Gomis, Jaume and Passerini, Filippo",
    title = "{Holographic Wilson Loops}",
    eprint = "hep-th/0604007",
    archivePrefix = "arXiv",
    reportNumber = "KUL-TF-06-11",
    doi = "10.1088/1126-6708/2006/08/074",
    journal = "JHEP",
    volume = "08",
    pages = "074",
    year = "2006"
}

@article{Gomis:2006im,
    author = "Gomis, Jaume and Passerini, Filippo",
    title = "{Wilson Loops as D3-Branes}",
    eprint = "hep-th/0612022",
    archivePrefix = "arXiv",
    doi = "10.1088/1126-6708/2007/01/097",
    journal = "JHEP",
    volume = "01",
    pages = "097",
    year = "2007"
}

@article{Drukker:2007dw,
    author = "Drukker, Nadav and Giombi, Simone and Ricci, Riccardo and Trancanelli, Diego",
    title = "{More supersymmetric Wilson loops}",
    eprint = "0704.2237",
    archivePrefix = "arXiv",
    primaryClass = "hep-th",
    reportNumber = "HU-EP-07-13, YITP-SB-07-12, IMPERIAL-TP-07-RR-02",
    doi = "10.1103/PhysRevD.76.107703",
    journal = "Phys. Rev. D",
    volume = "76",
    pages = "107703",
    year = "2007"
}

@article{Gomis:2008qa,
    author = "Gomis, Jaume and Matsuura, Shunji and Okuda, Takuya and Trancanelli, Diego",
    title = "{Wilson loop correlators at strong coupling: From matrices to bubbling geometries}",
    eprint = "0807.3330",
    archivePrefix = "arXiv",
    primaryClass = "hep-th",
    reportNumber = "NSF-KITP-08-96",
    doi = "10.1088/1126-6708/2008/08/068",
    journal = "JHEP",
    volume = "08",
    pages = "068",
    year = "2008"
}

@article{Aganagic:2002qg,
    author = "Aganagic, Mina and Marino, Marcos and Vafa, Cumrun",
    title = "{All loop topological string amplitudes from Chern-Simons theory}",
    eprint = "hep-th/0206164",
    archivePrefix = "arXiv",
    reportNumber = "HUTP-02-A024",
    doi = "10.1007/s00220-004-1067-x",
    journal = "Commun. Math. Phys.",
    volume = "247",
    pages = "467--512",
    year = "2004"
}

@article{Betzios:2021fnm,
    author = "Betzios, Panos and Kiritsis, Elias and Papadoulaki, Olga",
    title = "{Interacting systems and wormholes}",
    eprint = "2110.14655",
    archivePrefix = "arXiv",
    primaryClass = "hep-th",
    doi = "10.1007/JHEP02(2022)126",
    journal = "JHEP",
    volume = "02",
    pages = "126",
    year = "2022"
}

@article{Betzios:2023obs,
    author = "Betzios, Panos and Papadoulaki, Olga",
    title = "{Wilson loops and wormholes}",
    eprint = "2311.09289",
    archivePrefix = "arXiv",
    primaryClass = "hep-th",
    doi = "10.1007/JHEP03(2024)066",
    journal = "JHEP",
    volume = "03",
    pages = "066",
    year = "2024"
}

@article{Dvali:2000hr,
    author = "Dvali, G. R. and Gabadadze, Gregory and Porrati, Massimo",
    title = "{4-D gravity on a brane in 5-D Minkowski space}",
    eprint = "hep-th/0005016",
    archivePrefix = "arXiv",
    reportNumber = "NYU-TH-00-04-01",
    doi = "10.1016/S0370-2693(00)00669-9",
    journal = "Phys. Lett. B",
    volume = "485",
    pages = "208--214",
    year = "2000"
}

@article{Dong:2016fnf,
    author = "Dong, Xi",
    title = "{The Gravity Dual of Renyi Entropy}",
    eprint = "1601.06788",
    archivePrefix = "arXiv",
    primaryClass = "hep-th",
    reportNumber = "SU-ITP-16/01, SU-ITP-16-01",
    doi = "10.1038/ncomms12472",
    journal = "Nature Commun.",
    volume = "7",
    pages = "12472",
    year = "2016"
}

@article{PhysRevD.23.852,
  title = {Gravitational field of vacuum domain walls and strings},
  author = {Vilenkin, Alexander},
  journal = {Phys. Rev. D},
  volume = {23},
  issue = {4},
  pages = {852--857},
  numpages = {0},
  year = {1981},
  month = {Feb},
  publisher = {American Physical Society},
  doi = {10.1103/PhysRevD.23.852},
  url = {https://link.aps.org/doi/10.1103/PhysRevD.23.852}
}

@article{Maldacena:2001kr,
    author = "Maldacena, Juan Martin",
    title = "{Eternal black holes in anti-de Sitter}",
    eprint = "hep-th/0106112",
    archivePrefix = "arXiv",
    reportNumber = "NSF-ITP-01-59",
    doi = "10.1088/1126-6708/2003/04/021",
    journal = "JHEP",
    volume = "04",
    pages = "021",
    year = "2003"
}

@article{Tan:1991ay,
    author = "Tan, Chung-I",
    title = "{Generalized Penner models and multicritical behavior}",
    reportNumber = "BROWN-HET-810",
    doi = "10.1103/PhysRevD.45.2862",
    journal = "Phys. Rev. D",
    volume = "45",
    pages = "2862--2871",
    year = "1992"
}

@article{Penner:1986,
    author = "Penner, R. C.",
    title = "{The moduli space of a punctured surface and perturbative series}",
    journal = "Bull. Amer. Math. Soc.",
    volume = "15",
    pages = "73--77",
    year = "1986"
}

@article{Penner:1988cza,
    author = "Penner, R. C.",
    title = "{Perturbative series and the moduli space of Riemann surfaces}",
    journal = "J. Diff. Geom.",
    volume = "27",
    pages = "35--53",
    year = "1988"
}

@article{Harer1986,
    author = "Harer, J. and Zagier, D.",
    title = "{The Euler characteristic of the moduli space of curves}",
    doi = "10.1007/BF01390325",
    journal = "Inventiones mathematicae",
    volume = "85",
    pages = "457--485",
    year = "1986"
}

@article{berele1985hook,
  title={Hook flag characters and their combinatorics},
  author={Berele, A and Remmel, Jeff B},
  journal={Journal of Pure and Applied Algebra},
  volume={35},
  pages={225--245},
  year={1985},
  publisher={Elsevier}
}

@article{Lunin:2006xr,
    author = "Lunin, Oleg",
    title = "{On gravitational description of Wilson lines}",
    eprint = "hep-th/0604133",
    archivePrefix = "arXiv",
    doi = "10.1088/1126-6708/2006/06/026",
    journal = "JHEP",
    volume = "06",
    pages = "026",
    year = "2006"
}

@article{Lunin:2015hma,
    author = "Lunin, Oleg",
    title = "{Bubbling geometries for AdS$_{2}${\texttimes} S$^{2}$}",
    eprint = "1507.06670",
    archivePrefix = "arXiv",
    primaryClass = "hep-th",
    doi = "10.1007/JHEP10(2015)167",
    journal = "JHEP",
    volume = "10",
    pages = "167",
    year = "2015"
}
\bibliographystyle{utphys}

\end{document}